\documentclass[useAMS,usenatbib]{mn2e}
\usepackage[pdftex]{epsfig}
\usepackage{amsmath}
\usepackage{amssymb}
\usepackage{myaasmacros}
\usepackage[pdftex]{graphicx}

\def\deg{\hbox{$^{\circ}$} }
\def\HI{{\rm H\textsc{i}\ }}
\def\HII{{\rm H\textsc{ii}\ }}
\def\H2{\hbox{$\mathrm{H}_2$}}
\def\Gadget2{\rm{\textsc{Gadget\thinspace 2}\ }}
\def\kms{{\rm\thinspace km\thinspace s}^{-1}}
\def\cm{{\rm\thinspace cm}}

\def\kpc{{\rm\thinspace kpc}}

\def\Mpc{{\rm\thinspace Mpc\ }}   
\def\Msun{\hbox{$\thinspace M_{\odot}$}}
\def\pc{{\rm\thinspace pc}}     
       
\def\ergs{{\rm\thinspace erg\thinspace s}^{-1}}
       
\def\yr{{\rm\thinspace yr}}     
\def\Myr{{\rm\thinspace Myr}}     
\def\Gyr{{\rm\thinspace Gyr}}   
\def\Jy{{\rm\thinspace Jy}}

\title[Antennae simulations compared with {\em Herschel}-PACS]{Constrained
  simulations of the Antennae Galaxies: Comparison with {\em Herschel}-PACS
  observations\thanks{{\em Herschel} is an ESA space observatory with science instruments
    provided by European-led Principal Investigator consortia
    and with important participation from NASA.}}
\author[S. J. Karl, T. Lunttila, T. Naab, P. H. Johansson, U. Klaas,
M. Juvela]{S. J. Karl$^{1}$$^,$$^{2}$$^,$$^{3}$ \thanks{E-mail: skarl@ast.cam.ac.uk},
  T. Lunttila$^{4}$, T. Naab$^{3}$, P. H. Johansson$^{4}$,
  U. Klaas$^{5}$ and M. Juvela$^{4}$\\
$^{1}$ Institute of Astronomy, University of Cambridge,
Madingley Road, Cambridge CB3 0AH, United Kingdom\\
$^{2}$ Kavli Institute for Cosmology, Cambridge\\
$^{3}$ Max-Planck-Institut f\"ur Astrophysik,
  Karl-Schwarzschild-Str. 1, D-85741 Garching bei M\"unchen,
  Germany\\
$^{4}$ Department of Physics, University of Helsinki, Gustaf
H\"allstr\"omin katu 2a, FI-00014 Helsinki, Finland\\
$^{5}$ Max-Planck-Institut f\"ur Astronomie, K\"onigstuhl 17, D-69117
Heidelberg, Germany}
\begin{document}


\pagerange{\pageref{firstpage}--\pageref{lastpage}} \pubyear{2012}

\maketitle

\label{firstpage}

\begin{abstract}
We present a set of hydro-dynamical numerical simulations of the 
Antennae galaxies in order to understand the origin of the central overlap 
starburst. Our dynamical model provides a good match to the observed nuclear and 
overlap star formation, especially when using a range of rather inefficient
stellar feedback efficiencies ($0.01 \lesssim q_{\rm EoS}\lesssim
0.1$). In this case a simple conversion of local star formation to 
molecular hydrogen surface density motivated by observations accounts well
for the observed distribution of CO. Using radiative transfer
post-processing we model synthetic far-infrared spectral 
energy distributions (SEDs) and two-dimensional emission maps for
direct comparison with {\em Herschel}-PACS observations. 
For a gas-to-dust ratio of 62:1 and the best matching range of stellar
feedback efficiencies the synthetic far-infrared SEDs of the 
central star forming region peak at values of $\sim 65 - 81 \Jy$
at $99- 116 \mu$m, similar to a three-component modified black body
fit to infrared observations. Also the spatial distribution of
 the far-infrared emission 
at $70 \mu$m, $100 \mu$m, and $160 \mu$m compares well with the observations:
$\gtrsim50\%$ ($\gtrsim 35\%$) of the emission in each band is
concentrated in the overlap region while only $< 30 \%$ ($< 15 \%)$ is
distributed to the combined emission from the two galactic nuclei in
the simulations (observations). As a proof of principle we
show that parameter variations in the feedback model result in
unambiguous changes both in the global and in the spatially resolved
observable far-infrared properties of Antennae galaxy
models. Our results strengthen the importance of direct, spatially
resolved comparative studies of matched galaxy merger simulations as a
valuable tool to constrain the fundamental star formation and feedback
physics.
\end{abstract}


\section{Introduction}
\label{SF:intro}

Being one of the closest major mergers, NGC 4038/39 (Arp 244; also
known as ``the Antennae'') is an ideal laboratory for studying the details of
interaction-induced star formation in the local Universe. Especially
the overlap region between the two interacting disks is surprisingly
rich in molecular gas \citep{WilsonEtAl2000ApJ, GaoEtAl2001ApJ,
 2003ApJ...588..243Z} and is subject to strong large-scale shocks
driven by the galaxy collision \citep{HaasChiniKlaas2005A&A...433L..17H,
 2011A&A...534A.138H, 2012A&A...538L...9H, 2012ApJ...750..136W}. The 
spatial distribution of the molecular gas as traced by CO emission 
correlates well with the large population of young
massive star clusters in the Antennae \citep{WilsonEtAl2000ApJ,
 ZhangFallWhitmore2001ApJ, 2010AJ....140...75W, 2012ApJ...745...65U}. 

Consistently, early mid-infrared ISO observations showed that a
significant amount of the total star formation of the Antennae takes
place in the overlap region \citep{VigrouxEtAl1996A&A...315L..93V,
 MirabelEtAl1998A&A}. These results were confirmed by Spitzer
mid-infrared observations \citep{WangEtAl2004ApJS} and recent
{\em Herschel}-PACS far-infrared (FIR) data \citep{2010AA...518L..44K}. By combining
different bands in the FIR, \citet{2010AA...518L..44K} determined
local star formation rates (SFRs) in small, localized regions of the
Antennae. These observations showed that the Antennae galaxies harbour a
major off-nuclear starburst in the overlap region and an arc-like
star forming area in the northern galaxy. This makes the
Antennae an exceptional system, together with only a handful of other
mergers showing similar levels of off-nuclear interaction-induced star
formation, e.g. Arp 140 \citep{2007MNRAS.374.1185C}, II Zw 096 
\citep{2010AJ....140...63I}, NGC 6090 \citep{1999ApJ...525..702D,
 2004ApJ...616L..67W}, NGC 6240 \citep{1999ApJ...524..732T,
 2010A&A...524A..56E}, and NGC 2442 \citep{2010ApJ...723..530P}. This
behaviour contrasts with the general observed trend in a sample of 35
pre-merger galaxy pairs observed by \citet{2007AJ....133..791S} and with
results from previous numerical simulations 
\citep[e.g.][]{BarnesHernquist1996ApJ}. In the Antennae, the total SFR
in the overlap region even exceeds the combined SFR of the two
galactic nuclei by a factor of $\sim3$
\citep{2010AA...518L..44K}. While most of the
star forming regions across the Antennae show a modest level of star
formation, a few areas in the overlap region and the western loop are
sites of intense localized bursts with specific star formation rates
similar to those of heavily star-bursting ULIRGs
\citep{WangEtAl2004ApJS, 2010MNRAS.401.1839Z}.

Until recently, numerical models were not able to reproduce the
extraordinary star formation properties of the Antennae. They failed
to produce a sufficiently gas-rich overlap region and underestimated
the level of total star formation in the Antennae, which is measured
at a rate of $\sim 5-22 \Msun \yr^{-1}$ \citep[e.g.][]{StanfordEtAl1990ApJ,
 ZhangFallWhitmore2001ApJ, KniermanEtAl2003AJ....126.1227K,
 2009ApJ...699.1982B, 2010AA...518L..44K}. However, in the last 
few years there have been a number of promising numerical studies of
the Antennae, e.g. about the nature of the observed star formation and
the current interaction stage \citep{2008MNRAS.391L..98R,
  2010ApJ...715L..88K, 2010ApJ...720L.149T}, the interpretation of the
observed cluster age distribution \citep{2011ApJ...734...11K}, and the
generation and evolution of magnetic fields in this proto-typical
galaxy merger \citep{2010ApJ...716.1438K}. The nature of the overlap
starburst has been interpreted in two ways: (1) the gas-rich overlap
region might originate from the two progenitor disks being observed in
projection on the plane-of-the-sky \citep{Barnes1988ApJ}. The high
level of star formation in a spatially extended pattern may result
from efficient fragmentation and subsequent star formation
\citep{2010ApJ...720L.149T}; (2) the actively star forming overlap 
region may be a natural consequence of the special interaction phase
of the Antennae, shortly before the two galaxies merge, at a time when
both disks already overlap, interact and large-scale shocks induce
localised star formation \citep{2010MNRAS.401.1839Z,
 2010ApJ...715L..88K, 2010AJ....140...75W}.

\begin{figure}
\centering 
\includegraphics[width=0.5\textwidth]{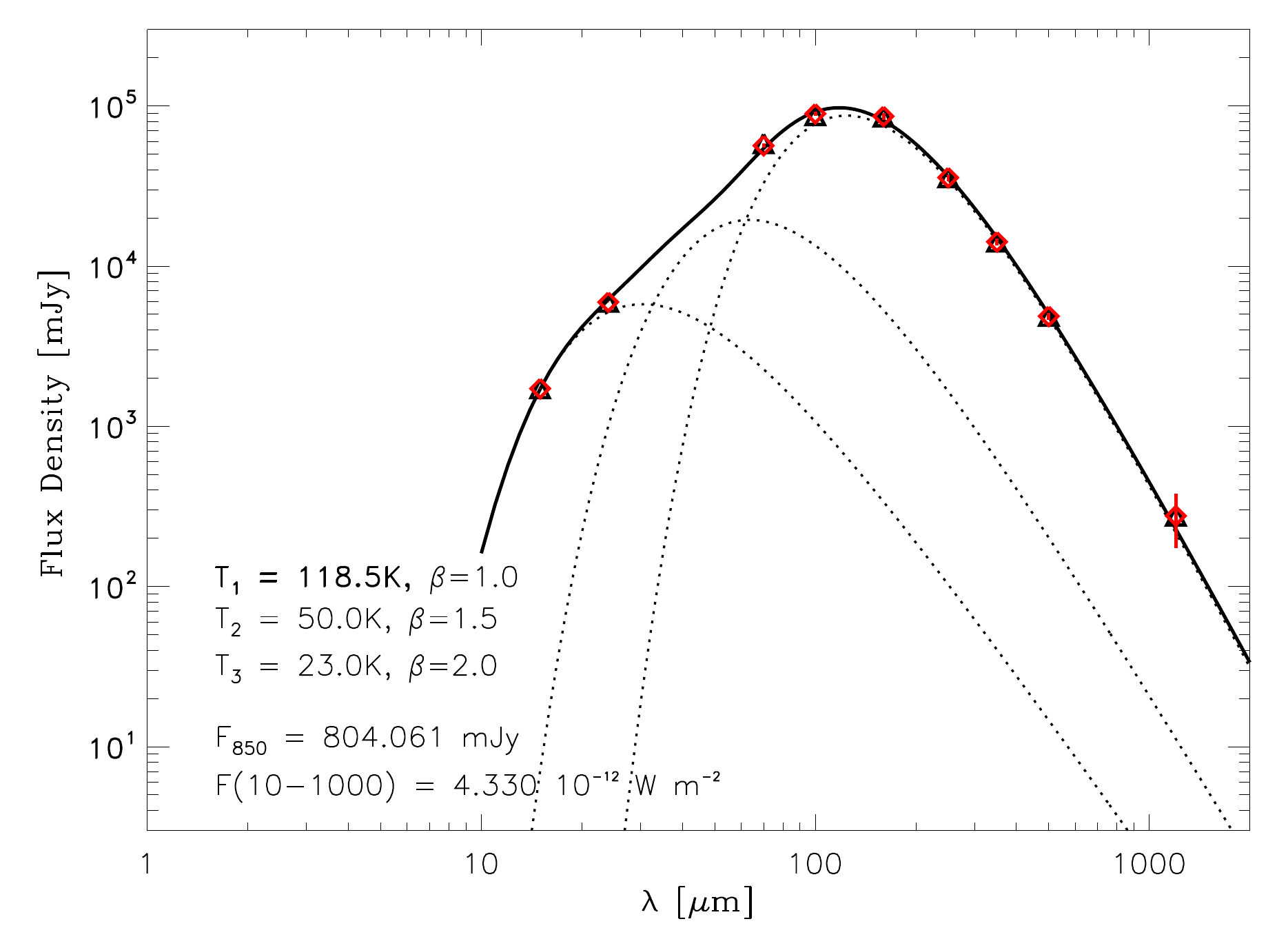}
\caption{Observed integral spectral energy distribution of the Antennae,
from ISOCAM 15\,$\mu$m, Spitzer-MIPS 24\,$\mu$m, {\em Herschel}-PACS
70, 100, and 160\,$\mu$m, {\em Herschel}-SPIRE 250, 350, and 500\,$\mu$m
and IRAM-MAMBO 1200\,$\mu$m photometry. Black triangles are the flux
values originating from the standard calibration of the respective
instrument. Red squares are colour corrected fluxes (colour correction 
only applied for the PACS and SPIRE photometry) according to the 
thermal components (dotted lines) fitted to result in a composite 
model SED (solid black line). Details of the modified black body fits 
are given in the figure legend as well as the 850\,$\mu$m flux 
prediction used for dust mass determination and the total 
10 -- 1000\,$\mu$m flux.}
\label{pic:ObsData}
\end{figure}

In this paper we re-visit the role of the physical properties in the
star forming interstellar medium (ISM) for the merger-induced star
formation in the Antennae. In particular, we aim to constrain the
net-effect of the applied thermal stellar feedback on the star
formation properties at the time of best match in a number of
simulations employing bracketing cases of the parameter of the
effective equation of state $q_{\rm EoS}$ in our adopted stellar
feedback algorithm. We find that the star formation histories in the
different simulations are complex and there is a large scatter in the
global observable star formation properties. Additional information
provided by a spatially resolved analysis of these properties
therefore seems to be needed to break degeneracies in simulations.
To this end, we post-process the simulation results of the Antennae
simulations, for the first time, using an efficient 3D Monte Carlo
radiative transfer method to compute the continuum spectral energy
distributions (SEDs) and to produce synthetic maps in different bands
in the far-infrared, which are then compared to detailed, spatially
resolved far-infrared observations of star formation sites in the
Antennae \citep[][see also Section \ref{SF:Obs}]{2010AA...518L..44K}.

We describe the merger simulations as well as details of the radiative
transfer post-processing procedure and the FIR observations in Section
\ref{SF:methods}. In Section \ref{SF:results}, we discuss the star
formation properties in the simulations subject to varying
efficiencies in the stellar feedback model. We compare to the FIR
observations in Section \ref{SF:CompareMaps}. Finally, we conclude
and discuss in Section \ref{SF:conclusion}.

\begin{figure*}
\centering 
\includegraphics[width=\textwidth]{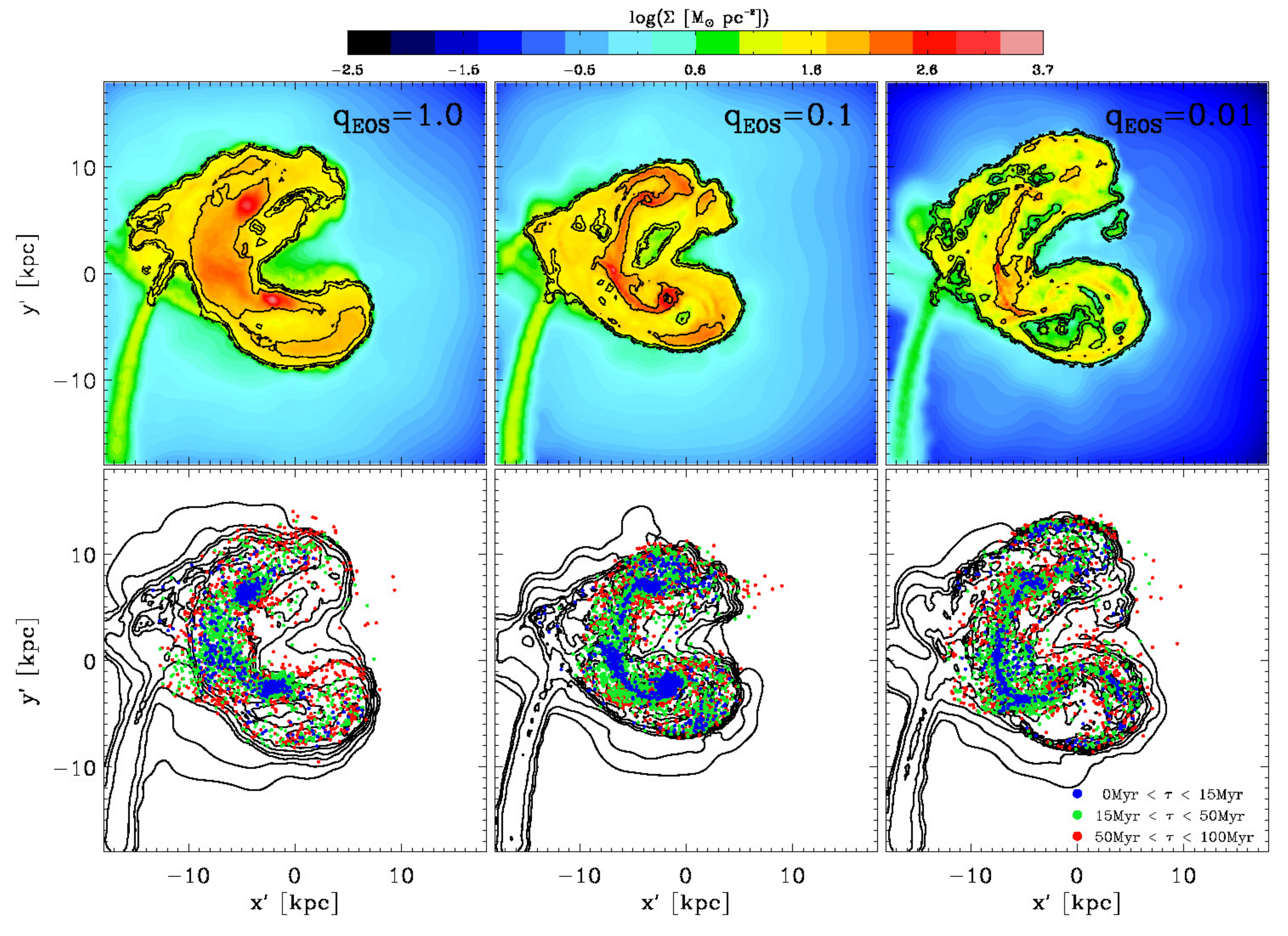}
\caption{Projected gas surface density and star formation in the
 central $\pm 18 \kpc$ of the galactic disks for the three Antennae simulations
 (K10-Q1, K10-Q0.1 and K10-Q0.01) with different values for the
 parameter for the softened equation of state, $q_{\rm EoS} = 1.0$ ({\it
  left}), $q_{\rm EoS} = 0.1$ ({\it middle}), and $q_{\rm EoS} =
 0.01$ ({\it right}). {\it Top panels:}
 Projected gas surface density overlaid with contours of the star
 formation rate surface density. The contours correspond to levels of
 $\log(\Sigma_{\rm SFR}\,/[\Msun \yr^{-1} \kpc^{-2}]) = -6,\,
 -5, \ldots, \,-3,\, -2$, respectively. {\it Bottom panels:} Gas
surface density contours and stellar particles formed within the last
$100 \Myr$, colour-coded by their age (see legend).}
\label{pic:SDs}
\end{figure*}

\section{Methodology}
\label{SF:methods}
\subsection{$N$-body$+$SPH simulations}
\label{SF:sims}
The numerical methods and simulations used in this study have, to
a major part, been presented in previous papers
(\citealp{2010ApJ...715L..88K, 2011ApJ...734...11K}, see also
\citealp{JohanssonEtAl2009ApJ}). Here, we briefly 
summarize the most important facts and describe the details of the new
simulation runs. All simulations were performed using the parallel
Tree/SPH code Gadget3 \citep[see][]{Springel2005MNRAS} including
cooling, star formation and thermal stellar feedback as detailed below.

Following \citet{KatzWeinbergHernquist1996ApJS}, radiative cooling is
computed for an optically thin plasma composed of primordial hydrogen and
helium, assuming ionization equilibrium in the presence of a spatially
uniform and time-independent local UV background 
\citep{HaardtMadau1996ApJ}.
Star formation and feedback are treated using the effective
multi-phase model by \citet{Springel&Hernquist2003MNRAS}, where stars
are allowed to form
on a characteristic timescale $t_\star$ in regions where gas densities
exceed a critical threshold density of $n_{\rm crit} = 0.128 \cm^{-3}$. In these
star forming regions, the interstellar medium (ISM) is assumed to
develop an effective two-phase fluid, where cold clouds are embedded
in a hot ambient medium in pressure equilibrium. This is achieved by
instantaneously returning mass and energy from
massive short-lived stars to the ISM upon formation of stellar
particles; we chose a mass fraction of $\beta = 0.1$ in massive
stars ($M_\star > 8 \Msun$) and $10^{51} \ergs$ of energy released per
supernova, according to a Salpeter-type stellar initial mass
function \citep{1955ApJ...121..161S}.

\begin{table*}
\caption{Nomenclature and discerning characteristics for the numerical
  simulations.}
\label{Tab:Models}      
\begin{minipage}{\textwidth}     
\centering          
\begin{tabular}{ c | c | c c c c | c | c | c}
\hline
\hline                             
Model\footnote{Note that simulations K10-Q0.01, K10-Q0.5, and K10-Q1 were
  already presented in \citet{2010ApJ...715L..88K} and
  \citet{2011ApJ...734...11K}.} & $M_\mathrm{vir}$\footnote{Masses
  refer to single galaxies in units of $10^{10}\Msun$, unless stated
  otherwise.} & $M_{\mathrm{disk, \star}}$ & $M_{\mathrm{disk, gas}}$
& $M_{\mathrm{bulge}}$ & $q_{\rm EoS}$ & $f_{\rm g}$ & RT eff. resolution\\
\hline                             
K10-Q0    & 55.2 & $3.3$ & $0.83$ & $1.4$ & 0    & 20\% & $2048^3$\\
K10-Q0.01 & 55.2 & $3.3$ & $0.83$ & $1.4$ & 0.01 & 20\% & $1024^3$\\
K10-Q0.1  & 55.2 & $3.3$ & $0.83$ & $1.4$ & 0.1  & 20\% & $2048^3$\\
K10-Q0.5  & 55.2 & $3.3$ & $0.83$ & $1.4$ & 0.5  & 20\% & $1024^3$\\
K10-Q1    & 55.2 & $3.3$ & $0.83$ & $1.4$ & 1    & 20\% & $1024^3$\\
\hline
\hline
\end{tabular}
\end{minipage}
\end{table*}

As a result of the two-phase formulation, the ``effective'' equation
of state (EoS) in star forming regions is quite ``stiff'', providing
strong and steeply rising pressure support to the gas with increasing
density (see Figure 1 in \citealp{Springel&Hernquist2003MNRAS}). A
convenient way of parametrizing uncertainties in the complex sub-grid
physics is to control the efficiency of the thermal feedback 
via a further dimensionless parameter, $q_{\rm EoS}$
\citep{SpringelDiMatteoHernquist2005MNRAS}, which linearly 
interpolates between a full, ``stiff'' feedback ($q_{\rm EoS} =
1$) and a softer, isothermal equation of state with $T=10^4$~K
($q_{\rm EoS} = 0$). The resulting equation for the pressure in star forming
regions then takes the form $P = q_{\rm EoS}\,P_{\rm eff} + (1-q_{\rm
  EoS})\,P_{\rm iso}$, where $P_{\rm eff} = (\gamma - 1)\, \rho \,
u_{\rm eff}$ and $P_{\rm iso} = (\gamma - 1)\, \rho \, u_{\rm
  iso}$. In this case, $\gamma = 5/3$ is the adiabatic index of an
ideal gas, $u_{\rm eff} = x\,u_{\rm cold} + (1-x)\,u_{\rm hot}$, the
effective specific internal energy weighting the
hot and cold phases by $x$, the mass fraction in cold clouds, and $u_{\rm iso}$ is the
specific internal energy 
corresponding to $T=10^4$~K. In this study, we test our Antennae
model with various choices for $q_{\rm EoS}$ against changes in the
quality of the match to {\em Herschel}-PACS observations in the FIR. 

Motivated by comparable total magnitudes in the B- and R-band for both
NGC 4038 and NGC 4039 \citep{1989spce.book.....L,
  1991trcb.book.....D}, we model the Antennae system 
as a 1:1 merger with a total mass of 
$M_{\rm tot} = 5.52 \times 10^{11} \Msun$ for each galaxy. 
The initial disk models are set up following the method described in
\citet{SpringelDiMatteoHernquist2005MNRAS}.
They all consist of a \citet{Hernquist1990ApJ} cold
dark matter halo with concentration parameter $c=15$, exponential
stellar and gaseous disks and a non-rotating Hernquist bulge with
masses $[M_{\rm halo}= 5.0\times 10^{11} \Msun, 
M_{\rm disk, tot}= 4.1 \times 10^{10} \Msun, M_{\rm bulge}= 1.4 \times
10^{10} \Msun]$, respectively. The disk is assumed to initially
contain a fraction $f_{\rm g} = 20\%$ in gas ($M_{\rm disk, \star} =
3.3 \times 10^{10} \Msun$, $M_{\rm disk, gas} = 8.3 \times 
10^{9} \Msun$). Here we present a total of five runs, each with varying
efficiencies of the stellar feedback using a softened equation of state
parameter with values of $q_{\rm EoS} = (0, 0.01, 0.1, 0.5,
1)$, as detailed in Table \ref{Tab:Models}.
We use identical parameters for both progenitor disks
except for the halo spin parameter $\lambda$ which directly
influences the disk scale length of the galaxies
\citep{MoMaoWhite1998MNRAS} and, hence, the length of the tidal tails
\citep{SpringelWhite1999MNRAS.307..162S}. Our choice of $\lambda_{\rm 
  4038} = 0.10$ and $\lambda_{\rm 4039} = 0.07$ results in disk 
scale lengths of $r_{\rm d,\, 4038} = 6.28 \kpc$ and $r_{\rm d,\, 4039}
= 4.12 \kpc$ (see \citealp{2010ApJ...715L..88K} for more details). 

At our standard resolution, each galaxy consists of a total of $N_{\rm
  tot} = 1.2 \times 10^6$ particles with $4\times 10^5$ dark matter particles,
$2\times 10^5$ bulge particles, and a total of $6\times10^5$ (stellar
and gaseous) disk particles. The particle numbers are
chosen such that we obtain a mass resolution of $m_{\rm bary} = 6.9
\times 10^4 \Msun$ with a gravitational softening length $\epsilon_{\rm bary} =
35 \pc$ in all baryonic components, while dark matter halo particles
have a mass of $1.2 \times 10^6 \Msun$ with $\epsilon_{\rm DM} = 150
\pc$. In the galactic disks, this choice yields $4.8\times10^5$
stellar and $1.2\times 10^5$ gas particles each.

The initial interaction orbit is set identical to the one presented in
\citet{2010ApJ...715L..88K}. The progenitors are set on a mildly
elliptic prograde orbit with initial separation $r_{\rm sep} = r_{\rm
  vir} = 168 \kpc$ and pericentric distance $r_{\rm p} = r_{\rm d,\,
  4038} + r_{\rm d,\, 4039} = 10.4 
\kpc$, where $r_{\rm vir}$ and $r_{\rm d}$ are the virial radius and
disk scale length, respectively. With disk inclinations $i_{\rm 4038}
= i_{\rm 4039} = 60\deg$ and pericentric arguments $\omega_{\rm 4038}
= 30\deg$ and $\omega_{\rm 4039} = 60\deg$ this orbit has proven to
result in a very good agreement with the observed large- and
small-scale morphology and line-of-sight kinematics of the Antennae at
the time of best match \citep[][]{2010ApJ...715L..88K}.

\subsection{Radiative transfer calculations}
\label{SF:RT}
In Section \ref{SF:results}, we compare synthetic FIR maps and SEDs
from our Antennae simulations with recent FIR observations by \citet{2010AA...518L..44K},
obtained with the {\em Herschel}-PACS instrument (see Section
\ref{SF:Obs}). We describe here the post-processing procedure applied
to our simulations to construct the synthetic FIR maps and SEDs.

\begin{table}
\caption{Model parameters of the radiative transfer calculations.}
\label{Tab:RTParameters}      
\centering          
\begin{tabular}{ c | c }
\hline
\hline                             
Property & RT parameter \\
\hline
effective resolution & $1024^3$ / $2048^3$\\                             
constant gas-to-dust ratio & 124\,:\,1 / 83\,:\,1 / 62\,:\,1\\
dust model & \citet{2003ARAA..41..241D}\\ 
stellar emission model & \citet{2003MNRAS.344.1000B}\\
stellar IMF & \citet{1955ApJ...121..161S}\\
metallicity  & $Z = 0.02$\\
initial disk particles ages & 0 - 6 Gyr\\
initial bulge particles ages & 5 - 9 Gyr\\
\hline
\hline
\end{tabular}
\end{table}

The FIR emission in star forming galaxies is dominated by
thermal emission from dust that has been heated by stellar
radiation. Due to strong spatial fluctuations in both the dust density
and the stellar emission, the dust temperature varies in a very
complicated manner. To produce meaningful FIR maps from the
simulations we model the three-dimensional radiative transfer (RT)
separately with a dust RT code \citep{2003A&A...397..201J,
  2005A&A...440..531J, 2012A&A...544A..52L}. Before starting the RT
calculations, the SPH snapshot is gridded onto an adaptive
mesh refinement (AMR) grid using the SPH smoothing kernel. The use of
an adaptive grid enables good spatial resolution in the dense inner
parts of the galaxy while keeping the total number of computational
cells in the domain low. For most of the simulations described in this
paper the size of the parent grid was chosen to be 80 kpc with an effective
resolution of $1024^3$, resulting in a minimum cell size of $\sim78
\pc$. For simulations K10-Q0 and K10-Q0.1, we have performed
additional tests with an effective grid resolution of 
$2048^3$ and minimum cell size of $\sim 39 \pc$ without
significantly changing the results. Details about the
resolution of the RT calculations are listed in Table
\ref{Tab:Models}. The total number of cells in the mesh refinement
grid is typically $\sim 5 \times 10^6$. Interstellar dust is assumed
to be uniformly mixed with the gas, and a single dust model is used
throughout the galaxy. As a starting point, we use the standard
($R_V = 3.1$) Milky Way dust model of
\citet{2003ARAA..41..241D}\footnote{http://www.astro.princeton.edu/$\sim$draine/dust/dustmix.html}. This
model implies a gas-to-dust 
mass ratio of 124:1. In addition, we test two models with a larger
dust mass. In these latter two models, the properties of the dust
grains are identical to the standard MW model, but the dust mass is
multiplied by a factor of $1.5$ and $2$, yielding gas-to-dust mass
ratios of $\approx 83:1$ and $62:1$, respectively. 
Stellar emission is modelled by assigning \citet{2003MNRAS.344.1000B}
SEDs to all star particles according to their ages using a
Salpeter stellar initial mass function \citep{1955ApJ...121..161S}. We
assume $Z=0.02$ \citep[``solar abundance''; ][]{1989GeCoA..53..197A}
for all stellar particles, i.e. disk, bulge, and newly formed stellar
particles, which is in reasonable agreement with estimates from young
star clusters in the Antennae \citep[][]{BastianEtAl2006A&A...445..471B,
  BastianEtAl2009ApJ...701..607B}. Stellar particles existing at the
start of the SPH simulation are assigned initial ages uniformly
distributed between 0 and 6 Gyr for the disk stars and between 5 and 9
Gyr for the bulge. Stellar particles formed over the course of the
simulation have their ages assigned corresponding to their formation
time. The choice for the particular disk and bulge stellar age
distributions has little influence on the FIR emission because the
dust heating is dominated by young stars, i.e., the stars formed in
the simulations, depending on their specific star formation histories.

After the dust distribution and radiation sources are set up, we
calculate the dust equilibrium temperatures, assuming that the gas
(and dust) distribution is smooth on the scale of individual grid
cells, i.e. we do not use any sub-grid model for \HII regions
\citep[c.f.][]{2010MNRAS.403...17J, 2011ApJ...743..159H}. All
calculations are run iteratively to account for the dust self-absorption
and heating until the dust temperatures have converged. 
Using the computed dust temperatures, we can construct SEDs in a
wavelength range of $50 \mu$m - $850 \mu$m and spatially resolved maps
in the 70 $\mu$m, 100 $\mu$m, and 160 $\mu$m passbands by integrating
the radiative transfer equation along the line-of-sight. The maps are
then convolved to the resolution of the {\em Herschel}-PACS instrument at
the corresponding wavelengths to simulate the observations by
\citet{2010AA...518L..44K}. The full-width-half-maximum (FWHM) of the
point-spread function (PSF) is $5.5\arcsec$, $6.8\arcsec$, and
$11.8\arcsec$ in the 70 $\mu$m, 100 $\mu$m, and, 160 $\mu$m bands,
respectively. For comparison with the observations, we re-sized the
simulated maps from the native RT resolution to the angular pixel
sizes used in the observations ($1.1 \arcsec$, $1.4 
\arcsec$ and $2.1 \arcsec$), assuming a distance $D=30.8 \Mpc$to the
Antennae. In addition, random Gaussian noise is added on top of the
simulated data at a level corresponding to the noise levels
in each spectral band, and we analyze the same spatial extent in the
simulated and observed maps. A summary of the RT model
parameters is given in Table \ref{Tab:RTParameters}.

\begin{table*}
\caption{Local star formation rates, gas masses, stellar mass
  formed within the last $15 \Myr$, and FIR fluxes in
  the galactic nuclei and the overlap region for the five Antennae
  simulations (K10-Q1, K10-Q0.5, K10-Q0.1, K10-Q0.01, and K10-Q0) with
  $q_{\rm EoS} = 1.0$, $q_{\rm EoS} = 0.5$, $q_{\rm EoS} = 0.1$,
  $q_{\rm EoS} = 0.01$, and $q_{\rm EoS} = 0.0$.} 
\label{Tab:EoS}      
\begin{minipage}{\textwidth}     
\centering     
\begin{tabular}{ c | c | c c c c}
\hline
\hline                             
Simulation & Region & SFR [$\Msun \yr^{-1}$] &  $M_{\rm gas}$
[$\times10^{9} \Msun$] & $M_\star(<15\Myr)$ [$\times10^7 \Msun$] &
$f_{\rm 70}$/$f_{\rm 100}$/$f_{\rm 160}$ [$\Jy$]\footnote{Fluxes at
  $70 \mu$m, $100 \mu$m, and $160 \mu$m, given in Jy, for models with
  gas-to-dust ratio $62:1$ and the observations (see Section
  \ref{SF:Obs}).}\\
\hline
 & Nuc$_{4038}$ & 1.71 & $1.60$ & 2.30 & 4.59 / 6.54 / 4.12\\
 $q_{\rm EoS} = 1.0$ & Nuc$_{4039}$ & 4.08 & $1.68$ & 6.30 & 12.6 / 17.1 / 8.01\\
 & Overlap & 0.49 & $2.58$ & 1.03 & 2.61 / 5.15 / 7.98\\
\hline
 & Nuc$_{4038}$ & $4.55$ & $1.49$ & 6.98 & 10.3 / 14.3 / 6.63\\
$q_{\rm EoS} = 0.5$& Nuc$_{4039}$ & $2.85$ & $1.15$ & 4.72 & 10.4 / 12.7 / 5.74\\
 & Overlap & $0.76$ & $3.05$ & 1.58 & 5.31 / 8.32 / 11.4\\
\hline
 & Nuc$_{4038}$ & 0.92 & $0.58$ & 1.48 & 4.48 / 5.24 / 2.81\\
$q_{\rm EoS} = 0.1$ & Nuc$_{4039}$ & 2.02 & $0.10$ & 3.39 &
9.91 / 11.8 / 5.65\\
 & Overlap & $26.09$ & $3.67$ & 16.23 & 22.5 / 39.4 / 35.3\\
\hline
 & Nuc$_{4038}$ & $0.49$ & $0.26$ & 0.75 & 3.03 / 3.29 / 1.53\\
$q_{\rm EoS} = 0.01$& Nuc$_{4039}$ & $0.44$ & $0.14$ & 0.55 &
3.67 / 3.76 / 1.59\\
 & Overlap & $10.88$ & $3.09$ & 10.22 & 29.7 / 39.3 / 27.6\\
\hline
 & Nuc$_{4038}$ & $0.99$ & $0.13$ & 1.48 & 4.16 / 3.92 / 1.47\\
$q_{\rm EoS} = 0.0$& Nuc$_{4039}$ & $0.46$ & $0.15$ & 0.70 & 3.31 / 3.44 / 1.35\\
 & Overlap & $2.24$ & $2.52$ & 3.04 & 8.44 / 11.4 / 10.1\\
\hline
\hline
 & Nuc$_{4038}$ & $1.57 \pm 0.04$ & $3.65$ & $-$ & $3.91 \pm 0.12$ / $6.73 
 \pm 0.20$ / $8.94 \pm 0.45$\\
Observations\footnote{SFRs are calculated from the total infrared
  luminosity, $L_{(50-1000)\mu{\rm m}}$, according to
  \citet{Kennicutt1998ApJ}, equation (3). FIR flux densities, gas
  masses, and SFRs for the overlap region are estimated as a lower
  limit by summing up the emission from knots $K1-K4$ (for D =
  $30.8\,$Mpc). See \citet{2010AA...518L..44K} for a definition of
  the different FIR knots. Observed gas masses denote molecular gas
  masses from \citet[][]{WilsonEtAl2000ApJ}, see
  \citealp{2010AA...518L..44K}, Table 2.} & Nuc$_{4039}$ & $0.72 \pm
0.02$ & $1.34$ & $-$ & $1.91 \pm 0.06$ / $3.53 \pm 0.11$ / $3.40 \pm 0.17$\\
& Overlap & $6.93 \pm 0.27$ & $6.16$ & $-$ & $21.6 \pm 0.4\ $ / $31.4 \pm
 0.7\ $ / $30.1 \pm 1.1$\\
\hline
\hline
\end{tabular}
\end{minipage}
\end{table*}

\subsection{{\em Herschel}-PACS observations of the Antennae}
\label{SF:Obs}

As observational reference data we use far-infrared maps at $70\,\mu$m,
$100\,\mu$m and 160\,$\mu$m \citep{2010AA...518L..44K} obtained with
the PACS instrument \citep{2010A&A...518L...2P} on-board the {\em Herschel}
Space Observatory \citep{2010A&A...518L...1P}. Here we do not use the
original maps published in \citet{2010AA...518L..44K}, but re-reduced
ones obtained in the context of a FIR study of a sample of nearby closely  
interacting galaxies \citep[in prep.]{Klaas2013}. The new processing
is done with the programme package Scanamorphos
\citep{2012arXiv1205.2576R}, resulting in an improved signal-to-noise
(S/N) ratio and hence somewhat fainter levels in the flux limits (by
factors of 1.84, 1.67 and 1.97, respectively). Furthermore, more accurate  
calibration factors for extended emission are applied which were
not yet available in early 2010 (see the PACS photometer point source
flux calibration
overview\footnote{http://herschel.esac.esa.int/twiki/pub/Public/\\ 
  PacsCalibrationWeb/pacs\_bolo\_fluxcal\_report\_v1.pdf \label{ftn}},
and the PACS photometer point spread function
report\footnote{http://herschel.esac.esa.int/twiki/pub/Public/\\
  PacsCalibrationWeb/bolopsf\_20.pdf}). 
The derived fluxes are thus systematically lower by $\sim 15\%$ than 
those published in \citet{2010AA...518L..44K}. The final photometric uncertainties 
of the new maps are dominated by the current PACS absolute flux calibration 
accuracy, which is 3\%, 3\%, and 5\% at $70\,\mu$m, $100\,\mu$m, and
$160\,\mu$m, respectively$^{\ref{ftn}}$.
In the 160\,$\mu$m filter, the [CII] 158\,$\mu$m line contributes a few percent
to the total flux as revealed by PACS spectra of the Antennae 
\citep{2011ApJ...728L...7G}. This means that the true 160\,$\mu$m continuum 
flux is likely lower by 1-2\% than the one quoted in Table 4. 

Together with the 15 and 24\,$\mu$m integral fluxes published in
\citet{2010AA...518L..44K}, $250\, \mu$m, $350\,\mu$m, and 500\,$\mu$m
photometry extracted from archival {\em Herschel} maps with the SPIRE
instrument \citep{2010A&A...518L...3G}, as well as 1.2\,mm 
photometry from an IRAM-MAMBO map, the PACS $70\,\mu$m, $100\,\mu$m,
and $160\,\mu$m photometry is used to constrain the 10 to 1000\,$\mu$m
spectral energy distribution of the Antennae galaxies as shown in
Figure \ref{pic:ObsData}. This SED is decomposed into three simple
modified black bodies (see Figure \ref{pic:ObsData}), which allows to
apply proper colour correction factors for the PACS and SPIRE
photometry. Furthermore, the sub-millimeter photometry enables an
accurate prediction of the unobscured 850\,$\mu$m flux which is used
to estimate the total dust mass according to formula (6) in
\citet{2001A&A...379..823K}. With a dust temperature of 23 K and an
emissivity index of $\beta$=2, this gives a dust mass M$_{\rm dust}$ =
$7.1 \times 10^{7}$\,M$_\odot$ (for D = $30.8\,$Mpc).

Similar to the IRAS FIR flux definition \citep{1988ApJS...68..151H,
  1985cgqo.book.....L} and its extension to the 170\,$\mu$m ISOPHOT
filter \citep{2000A&A...359..865S}, we use a PACS FIR flux parameter
for the wavelength range 59 to 206\,$\mu$m (effective band width of
the 3 PACS bands together), which is valid for the temperature range
20 - 80\,K and emissivity indices n = 0 - 2, being the ranges of
interest for galaxies. The conversion formula for the calculation of
the PACS FIR flux parameter from the PACS 70, 100 and 160\,$\mu$m
monochromatic flux densities will be presented in detail in
\citet[in prep.]{Klaas2013}. The total FIR luminosity is derived from
this FIR flux parameter in analogy to the IRAS far-infrared luminosity
calculation in \citet{SandersMirabel1996}. This is the input for the
calculation of observational IR star formation rates (SFR) according 
to formula (3) in \citet{Kennicutt1998ApJ}.

\section{Star formation in the Antennae simulations}
\label{SF:results}

 \subsection{Global star formation properties}

The top three panels of Figure \ref{pic:SDs} show colour-coded maps
of the total gas surface densities, $\mathrm{\Sigma}_{\rm gas}$,
indicating differences in the central morphologies obtained for three
Antennae simulations providing a varying degree of pressure 
support in the star forming gas by adopting three different equation of
state parameters (see Section \ref{SF:methods}), $q_{\rm EoS} = 1.0$,
$q_{\rm EoS} = 0.1$, and $q_{\rm EoS} = 0.01$ (from left to
right). Each gas surface density map is overlaid 
with contours of the SFR surface density $\mathrm{\Sigma}_{\rm
  SFR}$. For all three parameters, we find a similar large-scale
morphology: the nuclei of the progenitor disks are still well
separated, but connected by a ridge of high density gas. In the case
of very efficient feedback ($q_{\rm EoS}=1.0$; K10-Q1.0), a 
high-density overlap region forms, but the highest densities are found
in the galactic nuclei. In the simulations K10-Q0.1 and K10-Q0.01
with less vigorous feedback ($q_{\rm EoS}=0.1$ and $q_{\rm
  EoS}=0.01$), the high surface density regions tend to
become tighter and more pronounced in the overlap (middle and right
top panels). As a consequence we also find an increasing number of
low-density (``void'') regions inside the disks for lower values of
$q_{\rm EoS} \le 0.1$ due to the decreasing pressure support in the
star forming gas. The star formation contours directly support this
picture. Lowering the effective pressure, and hence the stellar
feedback, in star forming regions leads to more confined, localized
regions of intense star formation activity.
 In the bottom panels of Figure \ref{pic:SDs},
we show the corresponding contours of the gas surface density and
overplot all stellar particles (with individual masses of
$m_\mathrm{star} =  6.9 \times 10^4 M_{\odot}$), colour-coded
according to their ages, having formed in the last $\tau < 15 \Myr$
(blue), $15 \Myr < \tau < 50 \Myr$ (green), and $50 \Myr < \tau < 100
\Myr$ (red)\footnote{The corresponding stellar masses formed in the
  last $\tau < 15 \Myr$ are given in Table \ref{Tab:EoS} for different
regions in the merger.}. This period corresponds to the time span of the
interaction-induced rise in the star formation activity during the
recent second encounter. The youngest stars (blue) form predominantly
in regions of currently high gas densities, i.e. in the centres, the
overlap region, and in the arc-like features along the disks, very
similar to observations of the youngest clusters in the Antennae
\citep{WhitmoreEtAl1999AJ, 2010AJ....140...75W,
  WangEtAl2004ApJS}. This qualitatively confirms our earlier results
in \citet{2010ApJ...715L..88K}, indicating that our conclusions are
robust to altering the parameters of the adopted star formation
feedback algorithm. However, Figure \ref{pic:SDs} also shows that the star
formation in the galaxy centres is much more pronounced in run
K10-Q1.0 with $q_{\rm EoS} = 1.0$ (bottom left panel), while the
overlap starburst gains more relative importance for runs K10-Q0.1
($q_{\rm EoS} = 0.1$) and K10-Q0.01 ($q_{\rm EoS} =0.01$; see also
below). Furthermore, young stars ($\tau < 50 \Myr$) tend to be more
concentrated in the overlap region and the well-defined arcs in the
remnant spirals, while older stars ($50 \Myr < \tau < 100 \Myr$) are
dispersed more evenly throughout the disks. This captures a similar
trend observed in the Antennae \citep[e.g. ][their Figure
2]{ZhangFallWhitmore2001ApJ}. Therefore we conclude that, within a
range of feedback efficiencies $q_{\rm EoS} \lesssim 0.1$, the overlap
starburst seems to be driven by the most recent episode of star
formation in the Antennae, which was induced by the second close
encounter, while for values approaching the full stellar feedback
($q_{\rm EoS} \gtrsim 0.5$) the overlap starburst is significantly
suppressed.\\

\begin{table}
\caption{Global star formation rates, gas masses, and FIR fluxes,
  measured over the entire system, for all Antennae simulations
  (K10-Q1, K10-Q0.5, K10-Q0.1, K10-Q0.01, and K10-Q0) with varying
  feedback parameters: $q_{\rm EoS} = 1.0$, $q_{\rm EoS} = 0.5$,
  $q_{\rm EoS} = 0.1$, $q_{\rm EoS} = 0.01$, and $q_{\rm EoS} = 0.0$.} 
\label{Tab:Total}      
\begin{minipage}{0.48\textwidth} 
\centering     
\begin{tabular}{ c | c c c}
\hline
\hline                             
Simulation & SFR\footnote{Units for the numbers given in columns 2-4
  are in [$\Msun \yr^{-1}$], [$\times10^{9} \Msun$], and [$\Jy$],
  respectively.} & $M_{\rm gas}$ & $f_{\rm tot,\,70}$/$f_{\rm
  tot,\,100}$/$f_{\rm tot,\,160}$\footnote{Total fluxes at $70 \mu$m,
  $100 \mu$m, and $160 \mu$m, given in Jy, for models with gas-to-dust
  ratio $62:1$.}\\
\hline
$q_{\rm EoS} = 1.0$ & 7.16 & $13.2$ & 31.0 / 49.6 / 46.8\\
\hline
$q_{\rm EoS} = 0.5$ & 8.74 & $12.2$ & 34.0 / 51.7 / 45.1\\
\hline
$q_{\rm EoS} = 0.1$ & 30.1 & $11.8$ & 48.5 / 76.8 / 66.4\\
\hline
$q_{\rm EoS} = 0.01$& 14.3 & $9.28$ & 49.2 / 65.1 / 46.7\\
\hline
$q_{\rm EoS} = 0.0$ & 5.36 & $8.38$ & 26.8 / 34.0 / 26.7\\
\hline
\hline
Observations\footnote{For the observed SFRs and FIR fluxes we give
  values re-analyzed for this paper (see Section \ref{SF:Obs},
  \citealp{Klaas2013}, in prep.). Total gas masses in the galactic discs 
  for \H2 and \HI are adopted from \citet{GaoEtAl2001ApJ} and
  \citet{HibbardEtAl2001AJ}, assuming a distance of $D=30.8$~Mpc and a
  CO-to-\H2 conversion factor $\alpha_{\rm CO} = 0.8 \Msun \pc^{-2}\, (K\, \kms)^{-1}$
  \citep{1998ApJ...507..615D}.} & 
$19.1 \pm 1.4$ & $\sim9.6$ & 56.5 / 89.4 / 86.1\\ 
\hline
\hline
\end{tabular}
\end{minipage}
\end{table}

Thanks to the very good spatial resolution of the
{\em Herschel}-PACS instrument in the FIR, the previously determined IR SEDs
of individual star forming knots \citep[see
e.g.][]{2009ApJ...699.1982B} could be augmented by additional flux 
estimates at $70\, \mu$m, $100\, \mu$m, and $160\, \mu$m \citep[][in
prep.]{2010AA...518L..44K, Klaas2013}. The improved 
analysis described in Section \ref{SF:Obs} allows for a
precise determination of the total FIR luminosities and SFRs
in individual knots yielding integrated SFRs of $\mathrm{SFR}^{\rm
  4038} = (1.57 \pm 0.04) \Msun \yr^{-1}$ and $\mathrm{SFR}^{\rm 4039}
= (0.72 \pm 0.02) \Msun \yr^{-1}$ in the nuclei of NGC 4038 and NGC
4039, and $\mathrm{SFR}^{\rm OvL} \gtrsim 6.93 \Msun \yr^{-1}$, when
taking the sum over all knots in the overlap region as a lower limit
(i.e. knots ``K1'' to ``K4''; see Figure 3 in
\citealp{2010AA...518L..44K} for a definition).  
We compare these estimates to our simulated SFRs in the galactic
nuclei of NGC 4038 and NGC 4039 (defined as the central region of each
nucleus with a radius 
of $1 \kpc$), and the overlap region (chosen according to the central
morphology of the simulations). Details are provided in Table
\ref{Tab:EoS}, where we compare 
the total SFRs, the gas mass, the stellar mass formed in the last $15
\Myr$, and the FIR fluxes in the three PACS bands at
$70 \mu$m, $100 \mu$m, and $160 \mu$m for the different specific regions
in each simulation. 
In the case of efficient feedback ($q_{\rm EoS} = 1.0$ and $q_{\rm
  EoS} = 0.5$), the simulations shows stronger star formation activity
in the two nuclei compared 
to the overlap region. The simulations with low stellar feedback ($q_{\rm EoS} = 0.1$ and
$q_{\rm  EoS} = 0.01$), develop a very vigorous overlap starburst with
a SFR of $\sim26 \Msun \yr^{-1}$ and $\sim 11 \Msun \yr^{-1}$, which
is higher than the observed lower limit of $\gtrsim 6.93 \Msun
\yr^{-1}$ (by a factor of $\sim1.5-4$) together with a very low ratio of
$\mathrm{SFR}_\mathrm{nuclei}/\mathrm{SFR}_\mathrm{overlap} = 0.11$
and $0.09$ compared to $\lesssim0.33$ in the {\em Herschel}-PACS
observations. Simulation K10-Q0 without stellar feedback has the
lowest gas masses and SFRs of all simulations due to a very efficient
star formation and gas consumption earlier in the simulation (see
below).
The total instantaneous SFRs in the five simulations show a large scatter
within $5.4 \Msun \yr^{-1}\lesssim \mathrm{SFR} \lesssim 30 \Msun \yr^{-1}$, as
measured from the simulated SPH particles at the time of best match
(see Table \ref{Tab:Total}).
These values are in broad quantitative agreement with the range of
observed values between $\sim 5 - 20 \Msun \yr^{-1}$
\citep[e.g.][]{ZhangFallWhitmore2001ApJ} and with the value of $\mathrm{SFR} =
(19.1  \pm 1.4) \Msun \yr^{-1}$ we derive from the total IR
luminosity in this paper (see Table \ref{Tab:Total}).
Therefore, without attempting a
fine-tuned match, we conclude that adopting a rather inefficient
parametrization for the thermal stellar feedback, such as in simulations
K10-Q0.1 and K10-Q0.01, tends to give the most consistent results with the
observed spatial distribution of star forming regions in the Antennae
system.

\begin{figure*}
\centering 
\includegraphics[height=0.85\textheight]{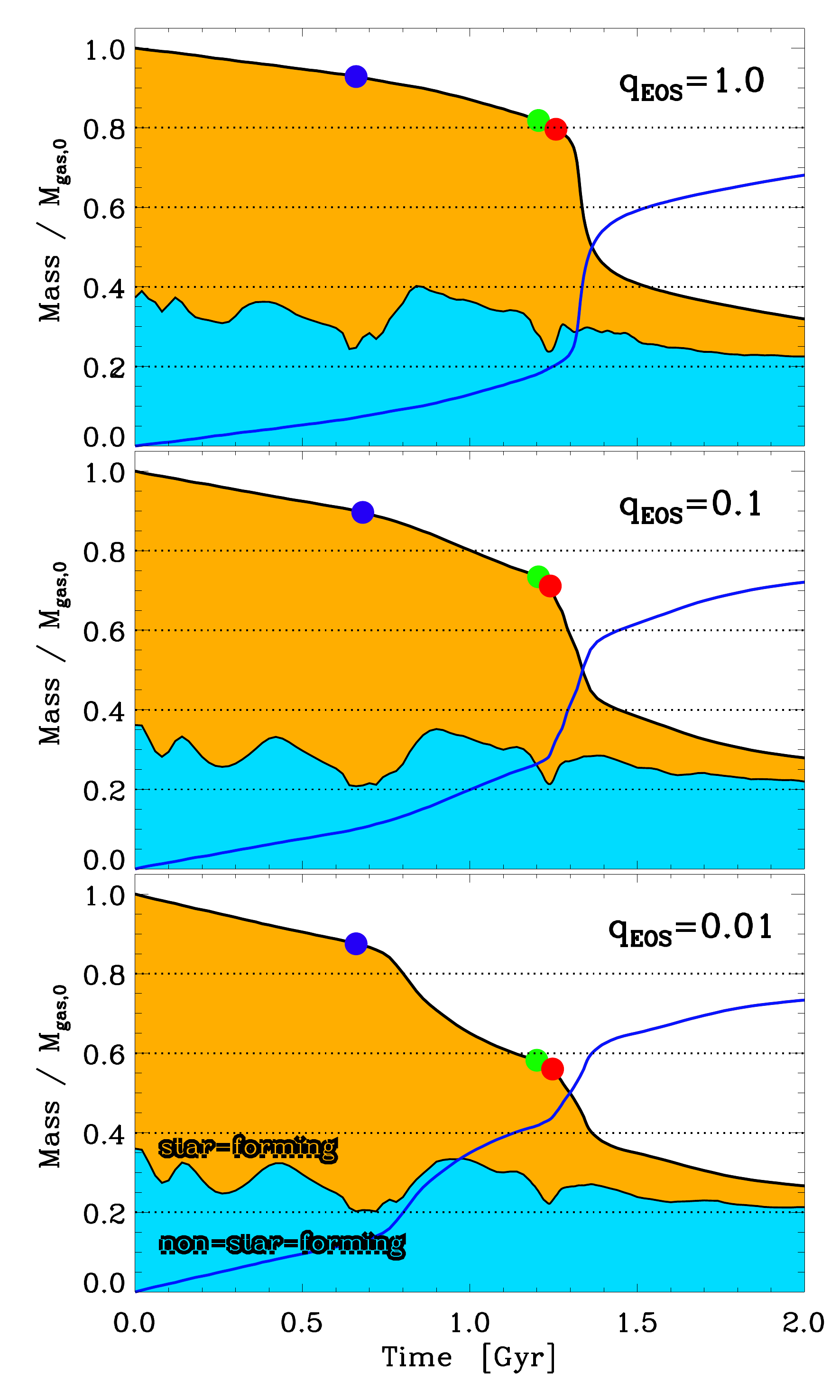}
\caption{Time evolution of the mass in gas (upper black solid line) and newly formed
  stars (blue line) normalized to the initial gas mass in three Antennae
  simulations with different parameters for the softened equation of state. {\it Top
    panel:}  full feedback simulation K10-Q1 with $q_{\rm EoS} = 1.0$; {\it middle
    panel:}  intermediate feedback simulation K10-Q0.1 with $q_{\rm EoS} = 0.1$; and {\it
    bottom panel:} the K10-Q0.01 simulation with very weak feedback, $q_{\rm EoS} = 0.01$. In
  addition, we distinguish between the dense star forming gas ($n > n_{\rm
    crit} \equiv 0.128 \cm^{-3}$) treated by the effective multi-phase model (orange) and
  non-star forming gas (blue). The blue, green and red dots indicate
  the times of first pericentre, second pericentre, and the
  time of best match in each simulation.} 
\label{pic:GasEvol}
\end{figure*}
%

\begin{figure*}
\centering 
\includegraphics[width=12cm]{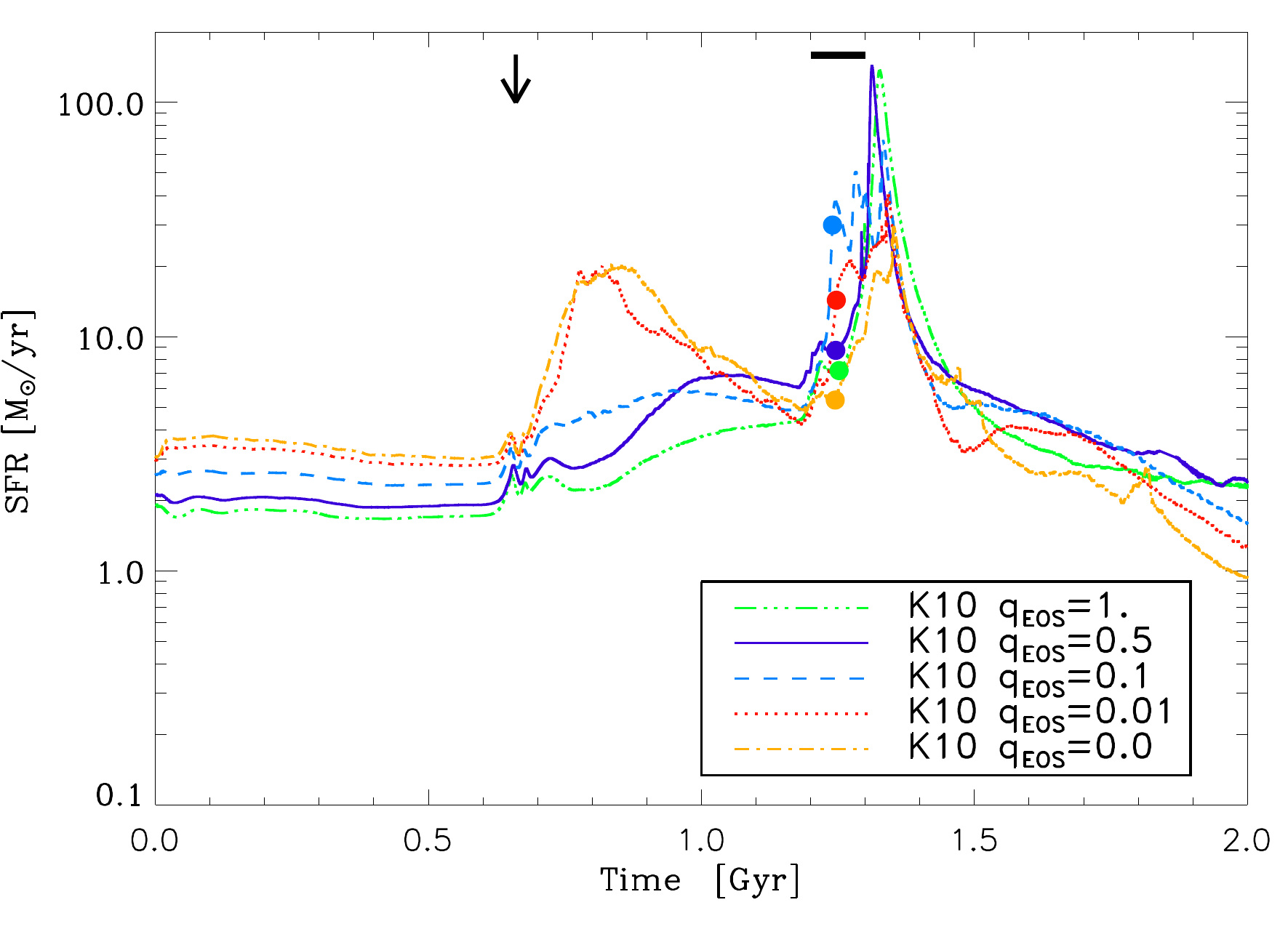}
\caption{Star formation histories for the five Antennae simulations
  with varying feedback parameters: $q_{\rm EoS} = 1.0$ (green),
  $q_{\rm EoS} = 0.5$ (purple), $q_{\rm EoS} = 0.1$ (blue), $q_{\rm
    EoS} = 0.01$ (red), and $q_{\rm EoS} = 0.0$ (orange). The arrow gives
  the time of first close encounter, while the bar indicates the time
  between second close encounter and final merger. The time of best
  match is denoted by the coloured dots for each simulation.}
\label{pic:SFHsEoS}
\end{figure*}

\subsection{Reproducing the overlap starburst}

In a first attempt to explain the large scatter in the star formation
properties, we show the evolution of the gas masses for three
simulations employing different parameters for the stellar
feedback model in Figure \ref{pic:GasEvol}. The gas is separated into
non-star-forming (blue) and star-forming gas phases (orange) based on
the density criterion $n > n_{\rm crit} = 0.128 \cm^{-3}$. All
three simulations show a steady decrease in their gas mass (upper solid
line) due to consumption by star formation from the very start of the
simulations, corresponding to a gradual build-up of the new stellar
component (blue solid line). However, due to the lower pressure
support in the effective multi-phase model in the K10-Q0.01 simulation
(bottom panel), a fraction of
$12.5\%$ of the initial gas mass has already been turned into
stars by the time of the first pericentre (blue dots), whereas the
K10-Q1.0 simulation, with a more efficient stellar feedback (top
panel), has only consumed about $7.2\%$ at the same time. The
K10-Q0.1 case is intermediate with $10.4\%$. The total fraction
of non-star-forming gas (blue) stays constant to within $\sim10 \%$ 
around a level of $0.3$ during the whole course of the simulated
time span ($t < 2.0 \Gyr$), save some little 'dips' during episodes of
induced star-formation associated with the first and the second
pericentre passages indicated by the blue ($t = 0.66 - 0.68 \Gyr$) and
green ($t \approx 1.20 \Gyr$) dots in Figure \ref{pic:GasEvol}. The 
more dramatic evolution can be seen in the star forming gas (orange,
$n > n_{\rm crit} = 0.128 \cm^{-3}$) which gets efficiently depleted
after the first pericentre in the K10-Q0.01 simulation with weak
feedback (bottom panel), having consumed more than $41\%$ of the
initial gas mass in the form of star forming gas at the time of second
pericentre. In contrast, in the full-feedback run (K10-Q1.0) there is
still about $81\%$ of the initial gas mass remaining after the second
encounter, which leads to a very rapid consumption of $40\%$ of
the gas within the first $\sim250 \Myr$ after the final merger (red
dots). The key difference is that the low feedback simulation
consumes the nuclear gas reservoir before the best match in contrast
with the full feedback simulation. In Table \ref{Tab:EoS} we
find that, with decreasing $q_{\rm EoS}$, the nuclear gas masses go
down drastically --- from nuclear gas masses of $M_{\rm gas} = 1.6
\times 10^9 \Msun$ for $q_{\rm EoS} = 1.0$ to  a factor of
$\sim10$ lower in the other two runs. The gas masses in the overlap
regions, on the other hand, show a less pronounced dependence with the
chosen effective stellar feedback parameter, staying constant to
within $\sim 40\%$, while the SFRs increase significantly for the two
lower-feedback simulations as discussed above.

A complementary view of this picture is offered by the star formation
histories in the Antennae simulations (Figure \ref{pic:SFHsEoS}).
The isothermal and quasi-isothermal simulations (K10-Q0 and K10-Q0.01),
show quite similar initial SFRs of $\sim 3.6 \Msun \yr^{-1}$ and $\sim
3.2 \Msun \yr^{-1}$. The
SFRs for the simulations with efficient feedback and $q_{\rm EoS} \ge 0.5$
(K10-Q0.5 and K10-Q1.0) are smaller by a factor of 
about $\sim 1.5$ from the very start, while run K10-Q0.1 is
intermediate with $\mathrm{SFR}(t=0) \sim 2.6 \Msun \yr^{-1}$.
After the first pericentre (black arrow) runs K10-Q0 and K10-Q0.01
both experience a significant rise in their SFRs by more than a factor of
five, while the other three simulations show a more gradual increase.
The low-feedback simulations show high total SFRs of $\sim30
\Msun \yr^{-1}$ (K10-Q0.1) and $\sim14 \Msun \yr^{-1}$ (K10-Q0.01)
at the time of best match (see Table \ref{Tab:Total}), powered by the
short-lived overlap starburst shortly after the second pericentre. The
full-feedback (K10-Q1.0) and intermediate-feedback (K10-Q0.5)
simulations, on the other hand, build up very powerful nuclear 
starbursts after the final merger with a maximum total SFR of
SFR$_{\rm max}\sim140 \Msun \yr^{-1}$ and SFRs at the best match of
only $\sim7 \Msun \yr^{-1}$ and $\sim9 \Msun \yr^{-1}$ (Table
\ref{Tab:Total}), respectively. An equally powerful starburst is
missing in the case of weaker stellar feedback with SFR$_{\rm max} =
68.4 \Msun \yr^{-1}$ for the K10-Q0.1, and SFR$_{\rm max} = 40.8 \Msun
\yr^{-1}$ for the K10-Q0.01 simulations. Having consumed about half of
its gas before the second pericentre, simulation K10-Q0 has the
lowest SFR of all simulations during the time between second pericentre and
final merger with a $\mathrm{SFR} \sim 5.4  \Msun \yr^{-1}$ at the best
match. Leaving aside simulation K10-Q0 without stellar feedback, we find a
clear anti-correlation between the height of the 
maximum SFRs after the first and second pericentres in the sense that
simulations having higher star formation peaks after the first
pericentre have correspondingly lower star formation peaks after final
coalescence and vice versa. Note also that this trend, interestingly,
yields very similar gas fractions in these simulations after final
coalescence of the galaxies (see Figure \ref{pic:GasEvol}), differing
by less than $\sim 5\%$ at 2 Gyr. It is, however, not directly
reflected in the SFRs at the time of best match, which we find shortly
($\sim 30-50 \Myr$) after the second encounter but before the time of
maximum star formation activity (see Figure \ref{pic:SFHsEoS}). This
is firstly due to a more rapid gas consumption during the first
pericentre in the weaker feedback runs and secondly due to a weaker
pressure support in the star forming gas phase in the K10-Q0.1 and
K10-Q0.01 simulations making the gas in the overlap region more
susceptible to local shock-induced fragmentation and subsequent star
formation, as discussed above. 

Overall, the star formation histories are complex and there is a large
scatter in the SFRs at the time of best match (being defined 
as the time when simulated and observed morphology and
line-of-sight kinematics are most similar), depending on the parameter
choice of $q_{\rm EoS}$ for the stellar feedback model. This, e.g.,
leads to very similar SFRs at the time of best match in the
full-feedback simulation K10-Q1, employing a vigorous feedback, and
the isothermal simulation K10-Q0 with no feedback at all. Therefore,
global properties like the total SFRs and the total gas masses alone
seem to be a degenerate diagnostic when comparing to observations.

\subsection{Comparison with observed CO maps}

\begin{figure*}
  \unitlength1cm
  \begin{picture}(14,7)
    \put(0,0){\epsfxsize = 7cm\epsfbox{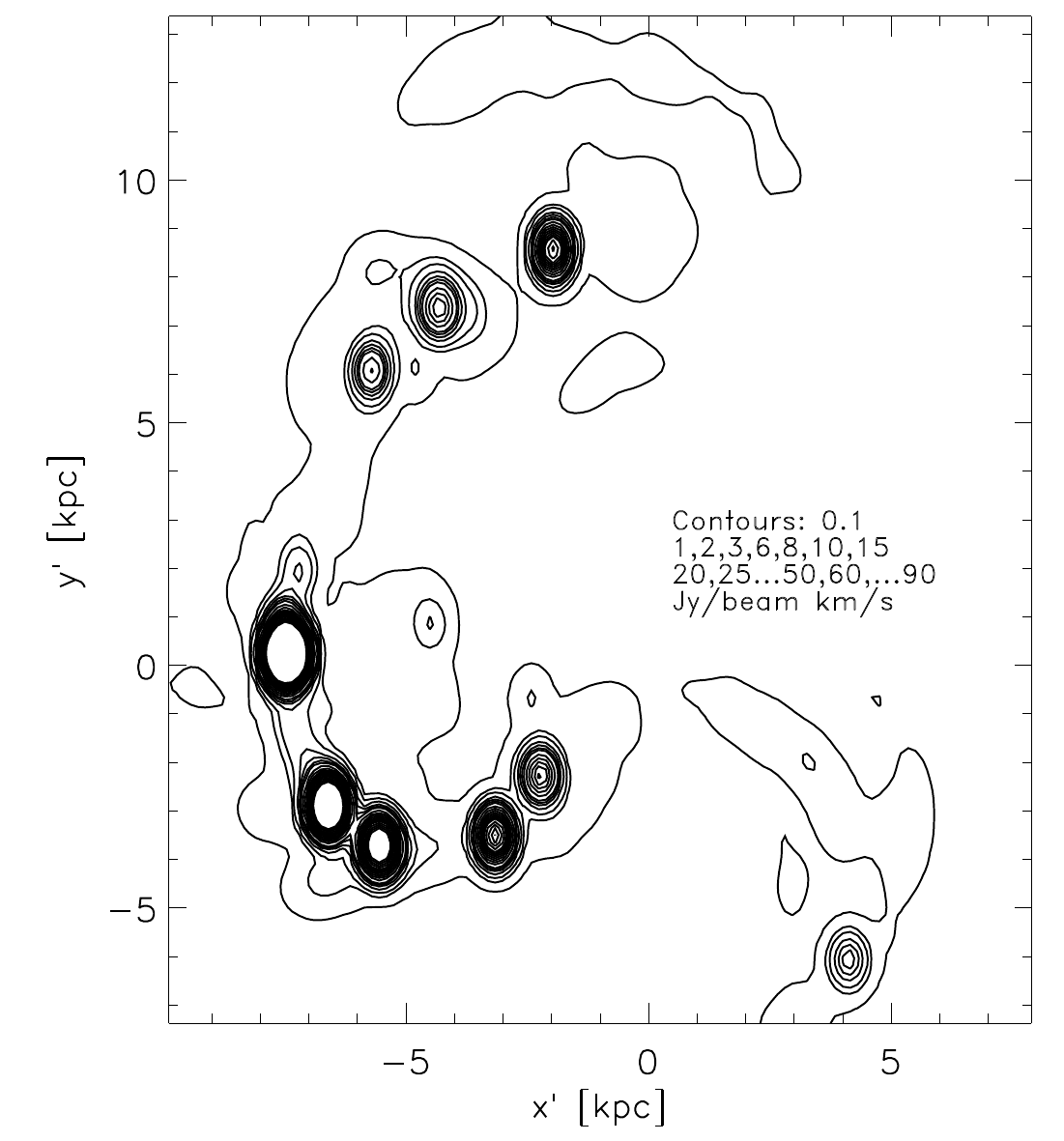}}
    \put(7,0.4){\epsfxsize = 7cm\epsfbox{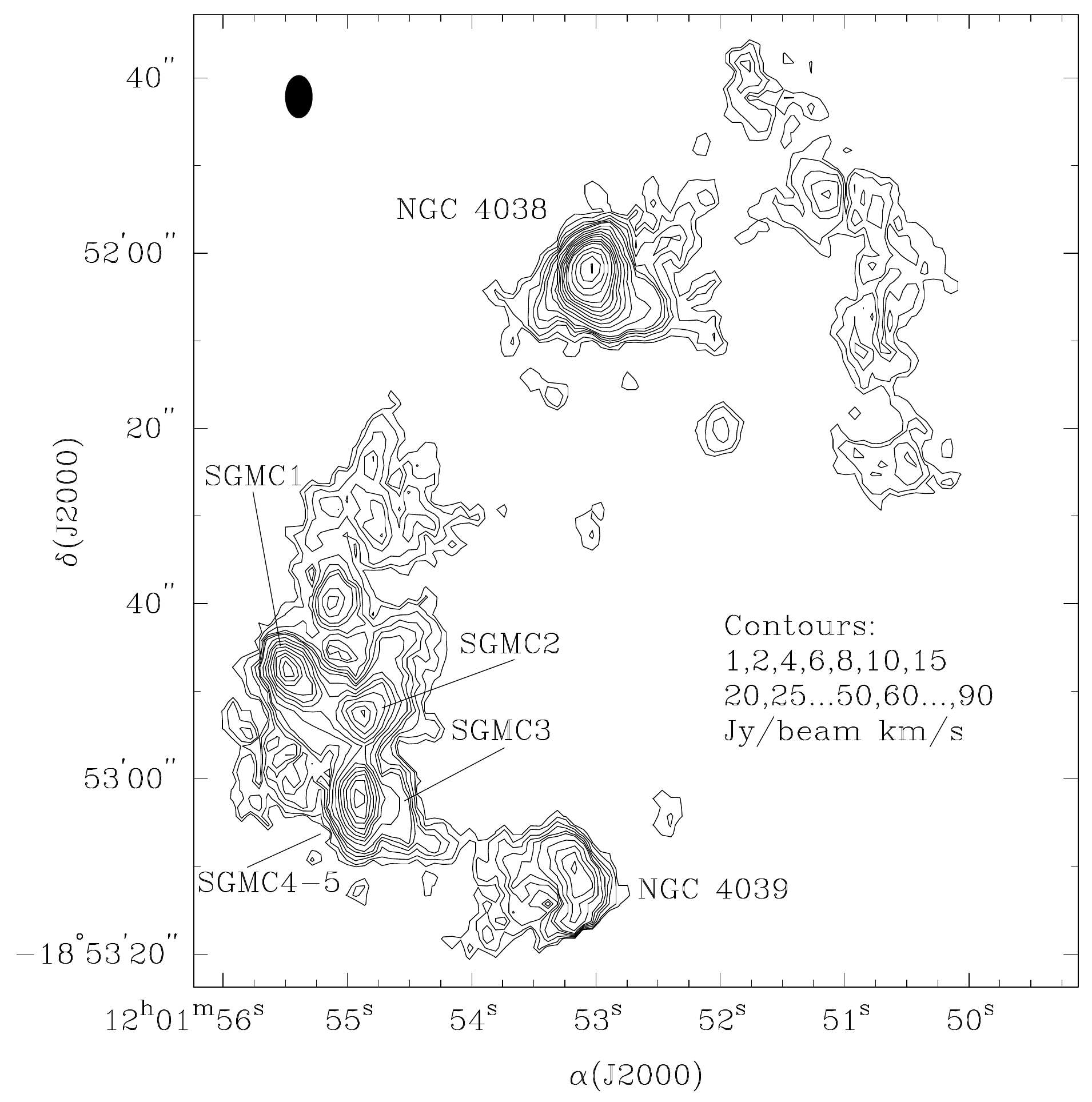}}
  \end{picture}
  \caption{{\it Left panel:} Synthetic CO map of the
    central regions of the Antennae from the K10-Q0.01
  simulation. The integrated CO masses are
  obtained in a two-step process. First, the \H2 surface densities are
  obtained from the SFR surface densities of the SPH particles using
  an ``inverse'' \citet{BigielEtAl2008AJ....136.2846B} relation. Then,
  adopting a conversion motivated by the observations
  \citep{WilsonEtAl2000ApJ}, we map \H2 masses back to CO integrated
  intensities (see text for details).
 {\it Right panel:} CO(1-0) integrated intensity map, adapted from Figure 1 in
  \citet{WilsonEtAl2000ApJ}. The two galactic nuclei and five super-giant
  molecular clouds in the overlap region are indicated. The filled
  black oval in the top left corner gives the size of the
  synthesized beam $(3\arcsec.15 \times 4\arcsec.91)$ in the observations.}
\label{pic:H2}
\end{figure*}

As a first reliability check of the hydrodynamical simulations,
we construct a simple CO-map derived from the simulated star formation
rates for comparison of the simulations to this widely used 
molecular gas tracer. Hereby, we first invoke the amount of molecular
\H2 from the SFRs of the simulated gas particles by using an
``inverse'' \citet{BigielEtAl2008AJ....136.2846B} relation of the form
\begin{equation} 
 \label{eq:Bigiel}
 \mathrm{\Sigma}_{\H2} = 10 \cdot \left(\frac{\mathrm{\Sigma}_{\rm SFR}}{A}\right)^{1/N} \Msun \pc^{-2},
\end{equation}
where $\mathrm{\Sigma}_{\H2}$ and $\mathrm{\Sigma}_{\rm SFR}$ are the surface densities in
\H2 and the SFR, respectively. The latter is given in units of
$\Msun \yr^{-1} \kpc^{-2}$, $A$
is the SFR surface density at a fiducial gas surface density of $10
\Msun \pc^{-2}$ with $\log(A[\Msun \yr^{-1} \kpc^{-2}]) = -2.06$, and
$N = 0.96$ is a power-law index near unity 
\citep[see][Equation 2 and Table 2]{BigielEtAl2008AJ....136.2846B}. 
To calculate the simulated SFR surface densities, we bin the gas
particles on a two-dimensional grid with pixel sizes corresponding to
$1\arcsec \times 1\arcsec$ at the assumed distance of $D =
30.8$~Mpc. Then, we relate the \H2 masses (per pixel)
back to integrated CO intensities (per pixel) by adopting the
conversion $M_{\rm mol} = 1.61 \times 10^4\, (D/\Mpc)^2\, S_{\rm CO} \Msun$
\citep{1990ApJ...363..435W, WilsonEtAl2000ApJ}, where $S_{\rm CO}$ is
given in $\Jy\, \kms$. Finally, we convert the map to units of $\Jy\,
\kms {\rm beam}^{-1}$ by multiplying the whole map with the beam area in
  pixels. The effective beam area is given as $1.133\times(3\arcsec.15 \times
4\arcsec.91)$ (FWHM) \citep{WilsonEtAl2000ApJ} where the factor
$1.133$ is a correction for the Gaussian shape of the beam\footnote{One should
bear in mind, however, that these maps still fundamentally represent the simulated
SFRs and no CO(1-0) emission, which we do not model in the simulations
presented here, even if we will call the converted simulated SFR maps 'CO-maps'
(instead of 'SFR-to-CO-maps') for simplicity for the rest of this paper.}.  
We show the synthetic CO map for simulation K10-Q0.01 in the left panel of
Figure \ref{pic:H2}, together with the observed integrated CO
map to the right (taken from Figure 1 in \citealp{WilsonEtAl2000ApJ}).
Here, we have applied a Gaussian filter to the simulated CO map,
corresponding to the FWHM sizes of the synthesized beam (see
above). We use the same contour range 
as the observations. However, since the simulated CO emission is more
concentrated (see also Section \ref{SF:CompareMaps} for a discussion)
we put an additional contour at $0.1$~Jy beam$^{-1}\,\kms$ to
account for the lack of extended emission in the simulation. The three
most luminous peaks in the simulation actually show higher peak values
(by up to a factor of five) than the highest contour level in the
observation. These can be seen as the blank spots within the three
peaks in the overlap region. In the synthetic CO map (left panel) we
find the CO emission to be mostly confined to the central regions of the merger
and a number of distinct high-density peaks in the 
southern overlap and near the two nuclei. To a lesser extent, i.e
on the additionally added, lowest level contour, we
also find emission from two arcs along the southern and northern remnant
arms, plus one prominent peak in the southern arm. The
observations exhibit a generally similar picture (right 
panel in Figure \ref{pic:H2}). The strongest emission peaks from the
CO(1-0) transition
are also detected in the two galactic nuclei and the overlap
region. However, as noted above, there is some discrepancy in
the more extended emission, in the sense that the north-western
spiral arc in NGC 4038 shows about $10$ times higher emission than
the simulation, and, there is no emission detected in the
south-western region of the southern disk as found at a low level and
in one distinct peak in the simulation. The apparent CO excess in the
southern simulated disk is qualitatively consistent with the high
total gas surface densities  in Figure \ref{pic:SDs}, as discussed
above and noted in \citet{2010ApJ...715L..88K}.

\section{Comparison with far-infrared observations}
\label{SF:CompareMaps}

\begin{figure*}
  \unitlength1cm
  \begin{picture}(15.6,22.0)
    \put(0,21){\epsfxsize = 15.6cm\epsfbox{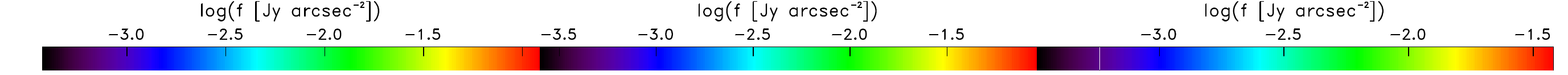}}
    \put(0,15.6){\epsfxsize = 15.6cm\epsfbox{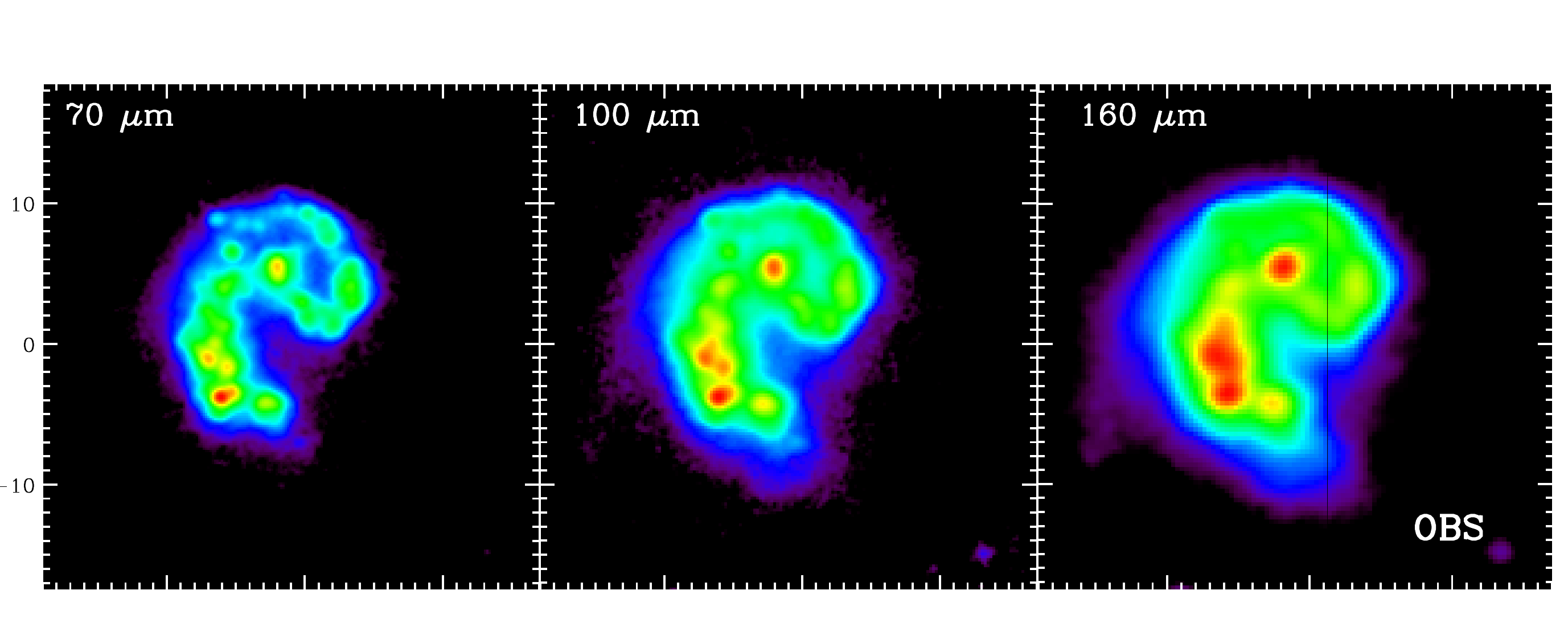}}
    \put(0,10.4){\epsfxsize = 15.6cm\epsfbox{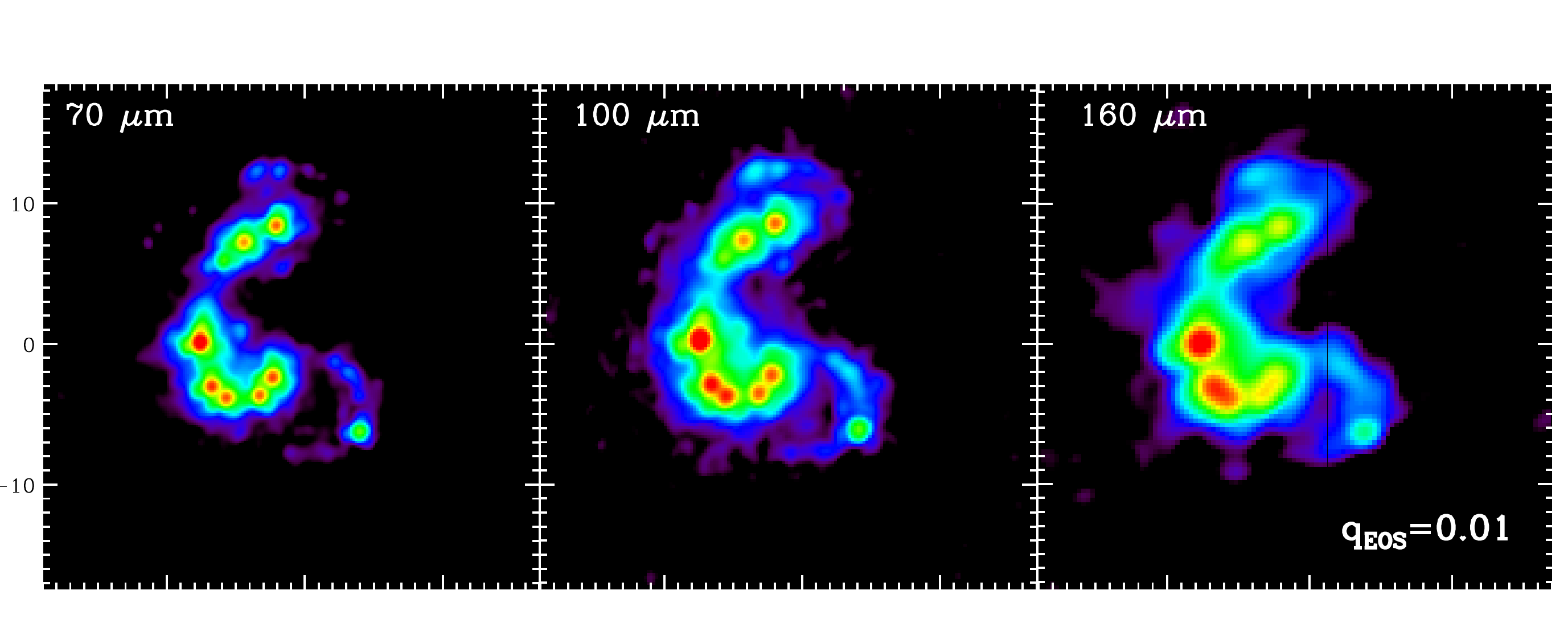}}
    \put(0,5.2){\epsfxsize = 15.6cm\epsfbox{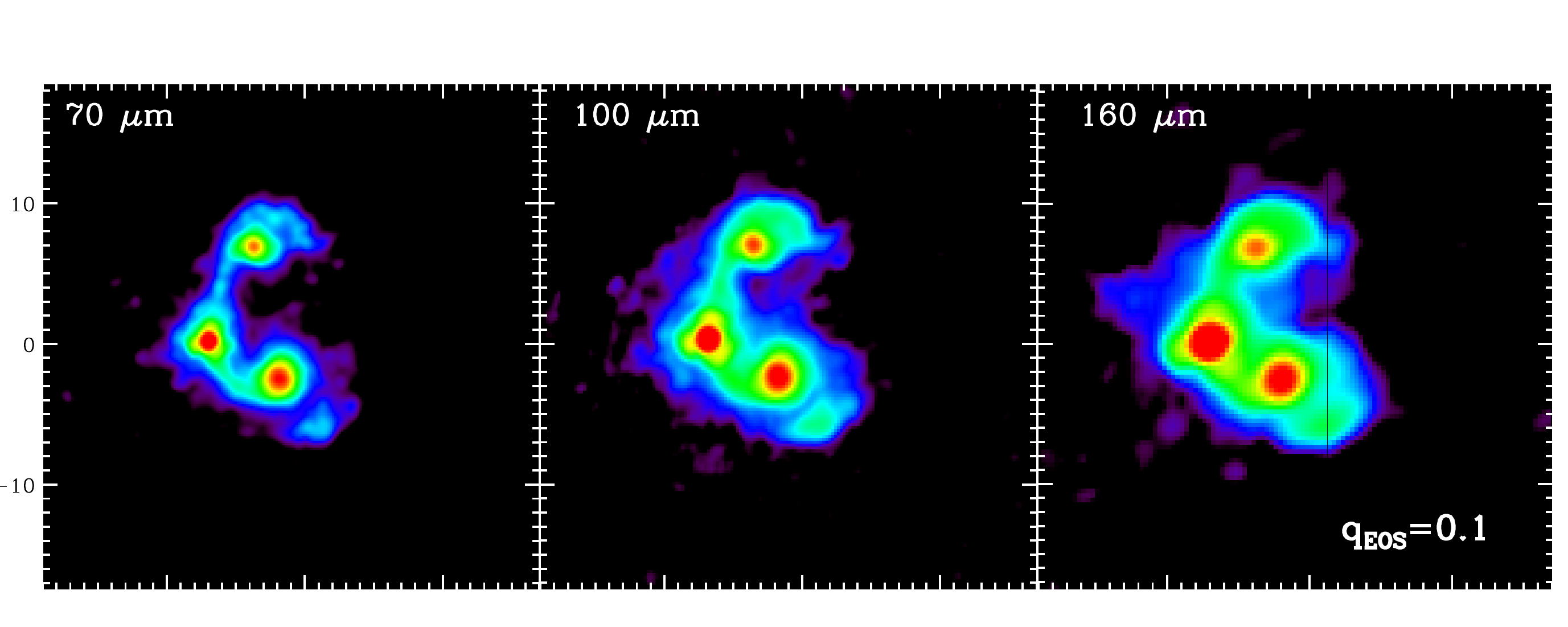}}
    \put(0,0){\epsfxsize = 15.6cm\epsfbox{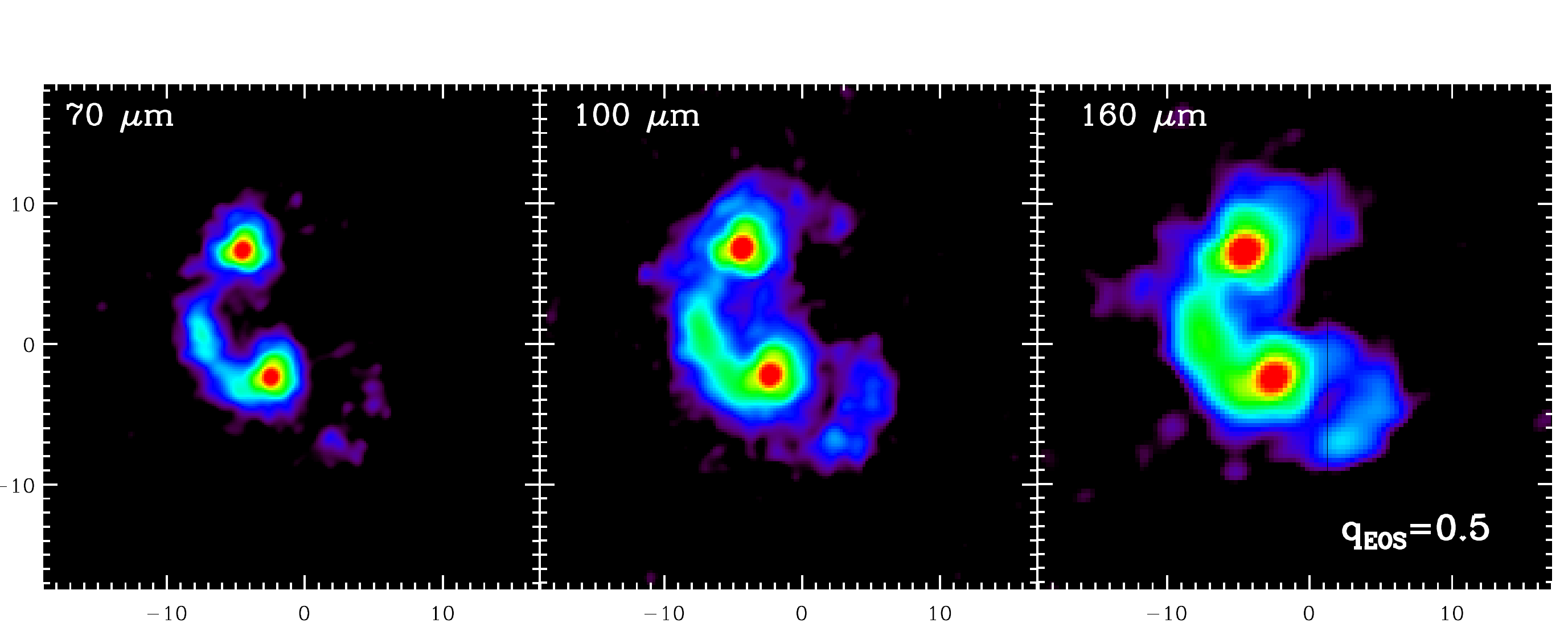}}
  \end{picture}
\caption{Synthetic 3D radiative transfer maps of the central
    regions of the Antennae at $70~\mu$m {\it (left column)},
    $100~\mu$m {\it (middle column)}, and $160~\mu$m {\it (right
      column)} for simulations K10-Q0.01, K10-Q0.1 and K10-Q0.5 with a
    gas-to-dust ratio $62:1$ {\it (bottom three rows)}. The simulated
    maps are compared to {\em Herschel}-PACS observations in the same bands
    \citep[{\it top row}; see Section
    \ref{SF:Obs}, ][in prep.]{2010AA...518L..44K, Klaas2013}. Surface
    brightness is given in logarithmic units of $\Jy/{\rm arcsec}^2$.}  
\label{pic:RT}
\end{figure*}

In Figure \ref{pic:RT} we compare {\em Herschel}-PACS data (top row; see
Section \ref{SF:Obs}) in three photometric wavebands centred on
$70\mu$m, $100\mu$m and $160\mu$m with the simulated FIR RT maps of
the K10-Q0.5, K10-Q0.1, and K10-Q0.01 simulations (bottom three rows; see
Section \ref{SF:RT}). Being an a priori un-constrained parameter in
the RT model (see Section \ref{SF:RT}), we tested for variations in
the gas-to-dust ratio (not shown) but found only negligible changes in
the FIR morphology for gas-to-dust ratios between $62 : 1$ and $124:1$
in all three FIR bands. Therefore, we only show the simulated
maps adopting a gas-to-dust ratio of $62:1$, which generally gives
the best match to the observed SED in the Antennae (see below). We
choose a lower cut in the surface brightness determined by the mean noise
level corrected for correlated noise above the mean background
in each band. The upper limits in the colour scheme are set by the
maximum surface brightness levels of the observations. 

Emission in the FIR bands reveals the sites of  recent star
formation 
that are still embedded in the dusty molecular 
clouds from which they have formed. The simulated FIR properties vary
significantly depending on the adopted efficiency of the
sub-grid stellar feedback model. Simulations K10-Q0.5 and
K10-Q0.1 (two bottom rows) show only a 
small number of emission peaks, most dominant either in the two nuclei
(K10-Q0.5) or in the overlap region (K10-Q0.1) in all three simulated bands. 
Simulation K10-Q0.01 (second row of Figure \ref{pic:RT}, on the other
hand, shows a well-defined pattern of 
sequential, confined star forming knots along the overlap
region very similar to the observations (top row). 
However, there are also two additional star-forming knots near
the northern nucleus  which are not detected at a similar
level in the observations. As seen in Tables \ref{Tab:EoS} and
\ref{Tab:Total}, more than $\gtrsim 50 \%$ of the FIR 
emission in each band originates from the overlap region in
simulations K10-Q0.01 and K10-Q0.1, compared to only $6-14 \%$ ($12-30
\%$) from the two nuclei in NGC 4038 and NGC 4039 combined in
K10-Q0.01 (K10-Q0.1). A similar percentage (10-15 \%) is observed in
the combined fluxes of the two nuclei in all three {\em Herschel}-PACS
bands. The combined emission from the observed knots in the
overlap regions gives a lower bound for the total emission from the
overlap region, yielding a total of $\sim35 - 39 \%$ for the three
bands. We find a very good agreement between the observed FIR surface
brightness maps and those of simulation K10-Q0.01

Apart from differences in the spatial distribution of
the simulated and observed FIR emission due to the specific adopted
stellar feedback parameters, there are also some characteristic
differences with respect to the observed FIR maps which are generic to
all the simulations. These are mostly caused by slight differences in
the underlying overall projected gas morphologies in the simulations, as
discussed in Section \ref{SF:results} (see Figure \ref{pic:SDs}).
Most notably,
the simulated RT maps in the bottom 
three rows in Figure \ref{pic:RT} indicate some low-level emission from
the southern disk (NGC 4039) which is not observed (top row).
This 
makes the observed disks look slightly more tilted to the north-west
than the simulated ones. 
Furthermore the dominant emission peak is found in the northern part
of the overlap region in the simulations which is slightly at odds
with the highest emission being observed in knots K1 and K2, 
at the southern end of the overlap region in the Antennae
\citep[Figure 3 in][]{2010AA...518L..44K}.
Comparing the global distribution of FIR emission, we do not find
equally strong spatially extended emission in the simulated progenitor
disks. This may be best seen in the northern spiral, e.g. comparing
the arc-like feature which is clearly detected in the $100~\mu$m and
$160~\mu$m {\em Herschel} maps (top middle and left panel) to the very
localized emission in the simulations in these bands (bottom three
middle and left panels). This may partly be caused by different
large-scale gas morphologies, as discussed above, and, to some extent,
by resolution effects and/or restrictions in the adopted
sub-grid model for the stellar feedback, leading to a spatially
confined star formation in high-density regions, which in turn is seen
in the simulated FIR maps. 

\begin{figure*}
  \unitlength1cm
  \begin{picture}(15.6,11.6)
    \put(0,10.6){\epsfxsize = 15.6cm\epsfbox{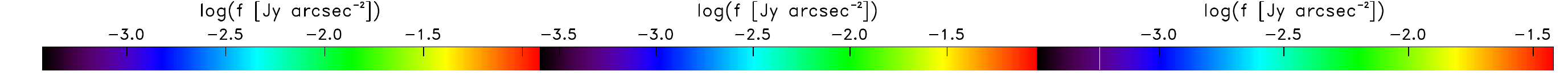}}
    \put(0,5.2){\epsfxsize = 15.6cm\epsfbox{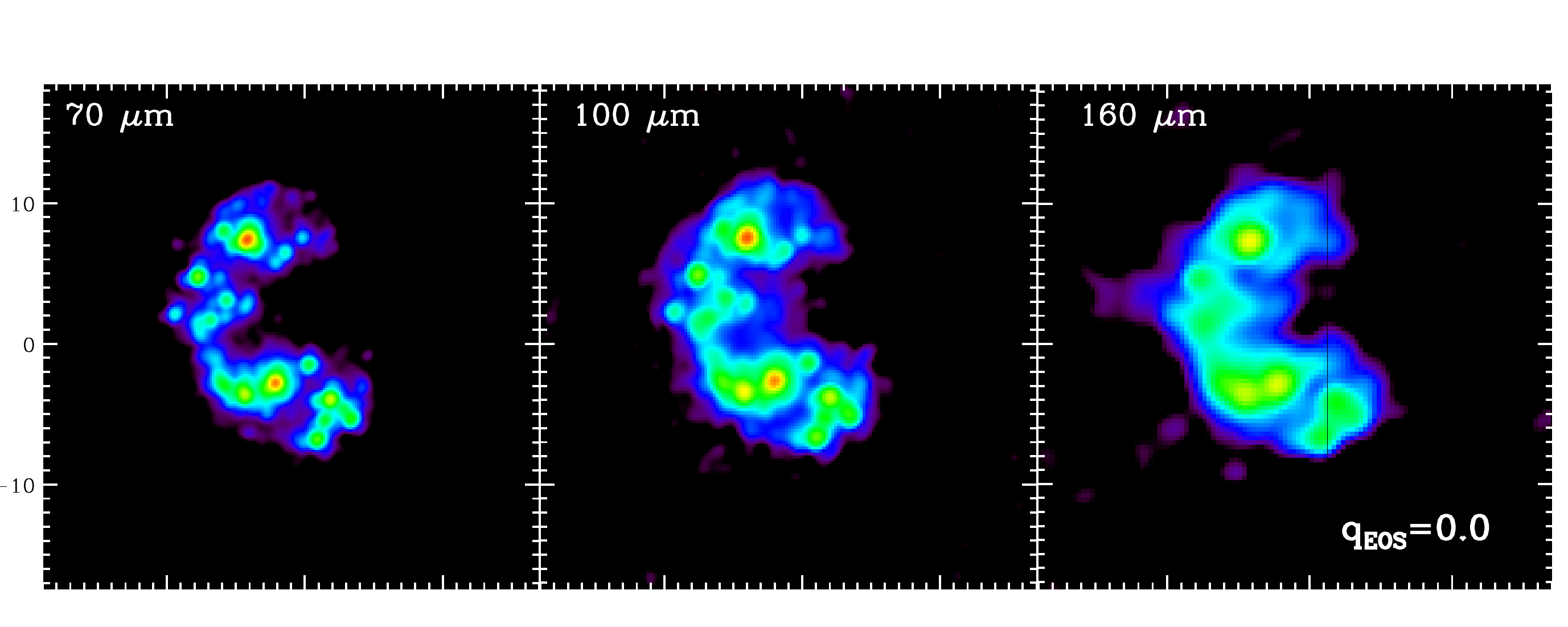}}
    \put(0,0){\epsfxsize = 15.6cm\epsfbox{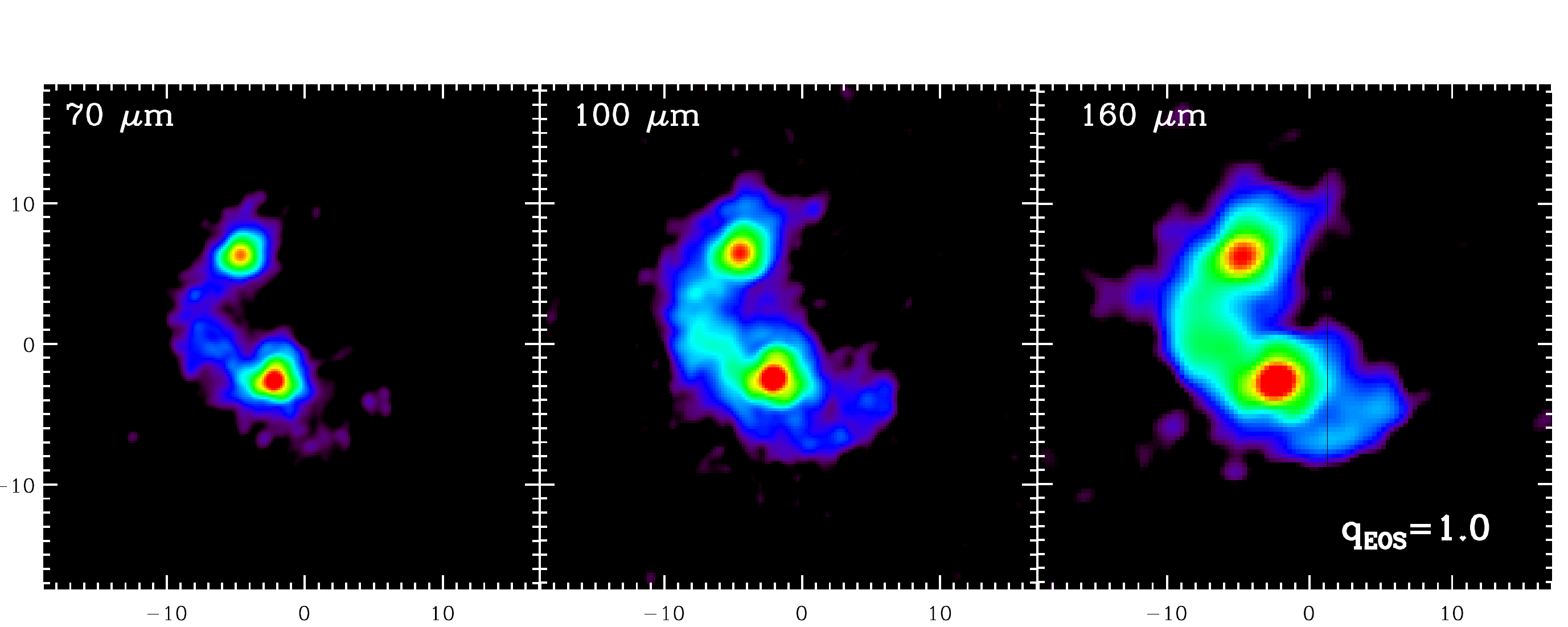}}
  \end{picture}
\caption{Synthetic 3D radiative transfer maps of the central
    regions of the Antennae at $70~\mu$m {\it (left column)},
    $100~\mu$m {\it (middle column)}, and $160~\mu$m {\it (right
      column)} with a gas-to-dust ratio $62:1$, highlighting the large
    scatter in FIR properties when adopting the two extreme parameter
    choices for the stellar feedback model in our sample: the
    simulation without stellar feedback K10-Q0 with $q_{\rm EoS}=0.0$
    {\it (top)} and the full-feedback simulation K10-Q1.0  with
    $q_{\rm EoS} = 1.0$ {\it (bottom)}. Surface brightnesses are given
    in logarithmic units of $\Jy/{\rm arcsec}^2$.}
\label{pic:RT2}
\end{figure*}

To highlight the striking differences in the simulated FIR properties
when adopting different feedback parameters we present in Figure
\ref{pic:RT2} the two most extreme cases from our sample, the
full-feedback simulation K10-Q1 with $q_{\rm EoS}=1$ and the isothermal
simulation K10-Q0 without any stellar feedback, i.e. $q_{\rm EoS}=0$, both
with a gas-to-dust ratio of $62:1$. In the  K10-Q1 simulation,
emission from the overlap region is feeble, especially at the 
shorter wavelengths, which reveals a prevalence of cold dust in this
region. The observed overlap starburst is missing entirely in
the synthetic FIR maps. In simulation K10-Q0 a large number of
spatially confined peaks of emission develops in the three FIR bands and the
emission shows a secondary peak at the southern end of the overlap
region. Due to the high prior gas consumption and a resulting low SFR  
of $\sim5.4 \Msun \yr^{-1}$ at the time of best match, however, the
surface brightness is much lower compared to the best-matching simulation
K10-Q0.01 (as discussed in Section \ref{SF:results}).

\begin{figure*}
\centering 
\includegraphics[width=0.8\textwidth]{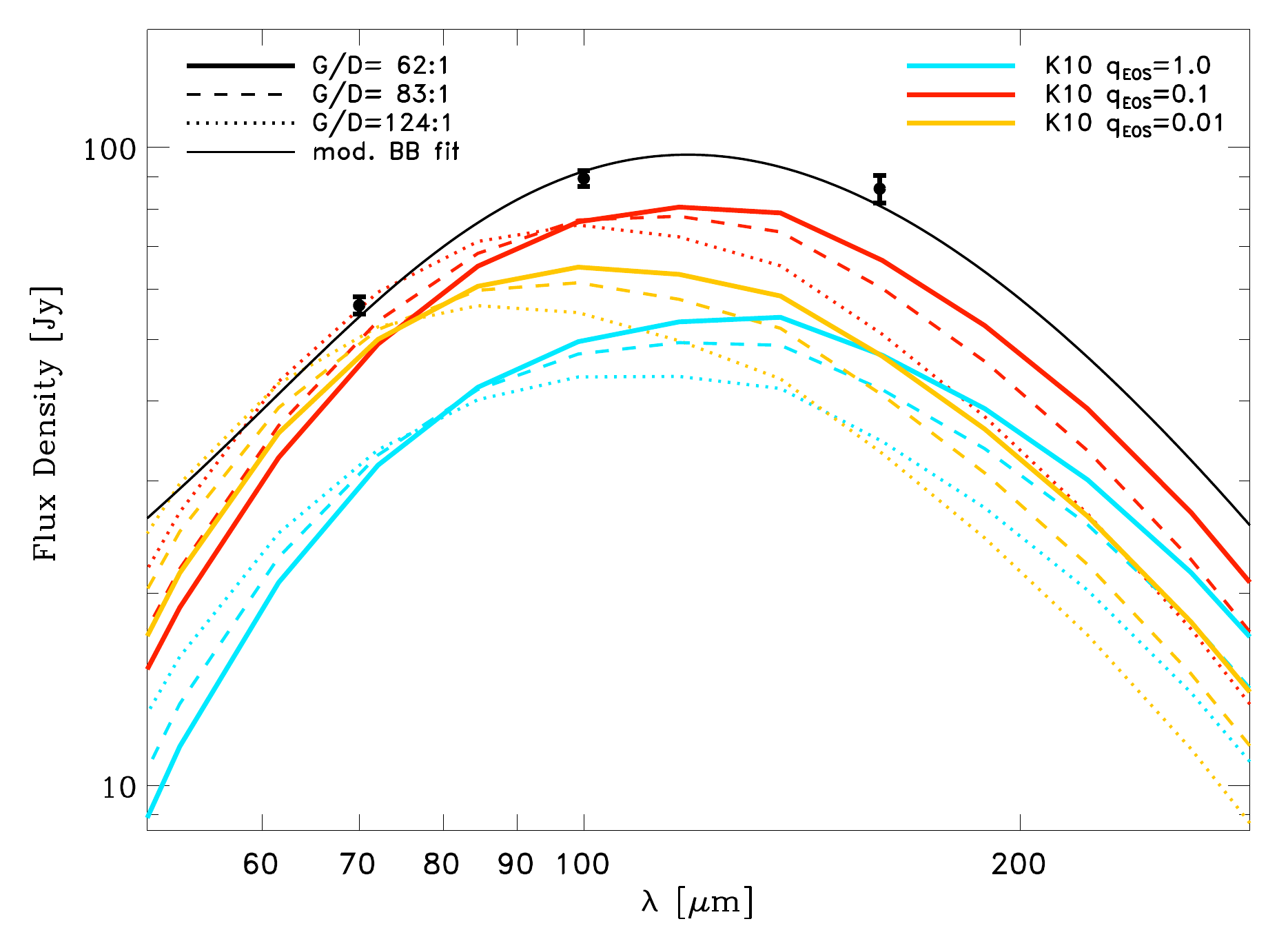}
\caption{Comparison of the total spectral energy distribution of
  radiative transfer calculations for the three different Antennae
  simulations K10-Q1, K10-Q0.1, and K10-Q0.01 with observations. Dotted
  lines correspond to SEDs calculated assuming a standard Milky Way
  gas-to-dust ratio of 124:1, while for the dashed and solid lines the
  ratio is reduced to a factor of $83:1$ and $62:1$,
  respectively. The small circles give the total fluxes obtained with
  {\em Herschel}-PACS observations at $70\,\mu$m, $100\,\mu$m, and
  $160\,\mu$m (see Figure \ref{pic:ObsData} and Section
  \ref{SF:Obs}, \citealp{2010AA...518L..44K, Klaas2013}, in prep.). The 
  total (intrinsic$+$systematic) photometric error, is given as bars,
  where we estimated the intrinsic photometric error as $\sim [50, 50,
  90]\,$mJy and the systemic error as $\sim [3\%, 3\%, 5\%]$ at
  $70\,\mu$m, $100\,\mu$m, and $160\,\mu$m, 
  respectively. The thin black line gives a theoretical fit to the
  observed IR SED using a three-component modified black body model
  (see Section \ref{SF:Obs}).}
\label{pic:SED}
\end{figure*}

Figure \ref{pic:SED} shows the simulated SEDs of the three Antennae
simulations with $q_{\rm EoS} = [1.0, 0.1, 0.01]$ in comparison to
observations of the FIR fluxes in the Antennae at 70$\mu$m, 100$\mu$m,
and $160 \mu$m, given as the filled dots with error bars, and a model of three
combined modified black body spectra fitted to IR data points within
$10-1000 \mu$m, given as the thin black line (see Figure
\ref{pic:ObsData} and Section \ref{SF:Obs}). The different 
lines for each simulation probe varying gas-to-dust models with a
standard Milky Way value of $124:1$ (dotted lines) and lower values
employing successively more dust with ratios 
of $83:1$ (dashed lines) and $62:1$ (solid lines). Increasing the
total amount of dust in the RT simulations leads to increasingly lower
dust temperatures seen as shifts in the peak emission towards longer
wavelengths. The observed spectral shape is best
reproduced with a low gas-to-dust ratio of $62:1$ (solid lines). 
In particular,
this choice leads to a steep slope between $70 \mu$m and $100 \mu$m,
similar to the observed data points, with a ratio
$f_{\mathrm{100}\mu\mathrm{m}}/f_{\mathrm{70}\mu\mathrm{m}} =
89.4/56.5 \approx 1.6$, as well as the flat and subsequently declining
part near the peak of the spectrum between $100 \mu$m and $160 \mu$m. 
Only for wavelengths below $\lesssim 60 \mu$m the simulated SEDs seem
to drop a bit too fast compared to the fitted modified  black body
spectra. This is in part due to the fact that, at these
and shorter wavelengths we start to miss some significant 
contributions to the total flux density from stochastically heated
small dust grains, or, in the mid-infrared, from polycyclic aromatic
hydrocarbons (PAHs). 

More important for the SEDs of the simulated data, however, is the
choice of the adopted stellar feedback parameter $q_{\rm EoS}$. While
the adopted gas-to-dust ratio changes mostly the spectral shape, the
parameter for the stellar feedback changes significantly both the
spectral shape and the total flux of the simulated SEDs due to
differences in geometry, total gas mass, and SFRs between the
simulations. Simulation K10-Q0.01 has the smallest amount
of gas (and dust) in the actively star forming regions (see
Table \ref{Tab:EoS}) combined 
with a relatively high SFR at the time of best match, yielding a strong
emission from hot dust at shorter wavelengths. Simulations K10-Q0.1 and
K10-Q1.0, however, show a spectral shape peaking at longer wavelengths
due to their larger reservoirs of gas (and dust)\footnote{Note here
  that average dust temperatures depend much more strongly on the
  total dust mass than on the SFR, or the bolometric luminosity
  $L_{\rm bol}$, causing the SED to appear ``warmer'' for smaller dust
  masses \citep[e.g.][]{2011ApJ...743..159H}.}.
The best agreement to 
the observed data points both in terms of the total flux level as well
as the spectral shape is reached for simulation K10-Q0.1 (i.e. with a
  low feedback efficiency of $q_{\rm EoS} = 0.1$) and a gas-to-dust ratio of 
$62:1$ (red solid line in Figure \ref{pic:SED}). The SEDs of simulations
K10-Q0.01 and K10-Q0.1 peak at a value of $\sim 65 \Jy$ at $99
\mu$m and $\sim81 \Jy$ at $116 \mu$m (compare also with Figure
\ref{pic:RT}), respectively, similar to the peak value of $97.4 \Jy$
at $118 \mu$m in the modified black body spectrum (thin black line).

\section{Summary \& discussion}
\label{SF:conclusion}

In this paper, we analyze the global and spatially resolved star
formation properties in a suite 
of hydro-dynamical simulations of the Antennae galaxies in order to
understand the origin of the observed central overlap starburst. Our
aim is to assess the
impact of variations in the applied thermal stellar feedback on
observable simulated star formation properties at the time of best
match, in the spirit of a study on the Mice galaxies by
\citet{Barnes2004MNRAS} who investigated variations due to the applied
star formation law. 
Using a state-of-the-art dust radiative transfer code
\citep[][see Section \ref{SF:RT}]{2012A&A...544A..52L} we
directly compare synthetic FIR maps and SEDs with re-analyzed
{\em Herschel}-PACS observations \citep[][see Section \ref{SF:Obs};
\citealp{Klaas2013}, in prep.]{2010AA...518L..44K}. This is the first
time that the FIR properties of a specific local merger system are
compared in quantitative detail to the simulated FIR properties of a
dynamical merger model.

The distributed extra-nuclear starburst in the
Antennae galaxies is plausibly a natural
consequence of induced star formation in high density regions in the
overlapping galactic disks after the recent second close
encounter. This is
the only phase in the simulations when a 
gas-rich, star-bursting overlap region forms and when we find
spatially confined areas of high star formation activity and of
similar extent as observed in the overlap region in the Antennae 
\citep[e.g.][]{MirabelEtAl1998A&A, ZhangFallWhitmore2001ApJ,
  WilsonEtAl2000ApJ, WangEtAl2004ApJS, 2010MNRAS.401.1839Z}. We
confirm that this result holds for a range of adopted stellar feedback
efficiencies of the sub-grid star formation model (see e.g. Figure 
\ref{pic:SDs}). However, we find a trend that the
overlap star formation activity is significantly suppressed for
values approaching the full stellar feedback simulation ($q_{\rm
  EoS} \lesssim 1$; \citealp{Springel&Hernquist2003MNRAS}). Best
qualitative agreement with observations is obtained using a rather
inefficient parametrization, $0.01 \lesssim q_{\rm EoS} \lesssim 0.1$, for the
thermal stellar feedback. This is in agreement with recent results
by \citet{2010MNRAS.407.1529H} who advocated a range of $q_{\rm
  EoS} = [0.125,0.3]$ by comparing model predictions and
simulations for different values of $q_{\rm EoS}$ to the gas
properties in observed star forming systems.

The star formation histories in the different simulations are
complex and there is a large scatter in the global observable
properties, e.g. the total gas mass, star formation rate and FIR
fluxes, at the time of best match (see Table \ref{Tab:Total};
Figures \ref{pic:GasEvol} and \ref{pic:SFHsEoS}), even though they
only differ in the chosen parameter for the thermal stellar
feedback. We find that additional information provided by a
spatially resolved analysis of these properties is needed to break
degeneracies in the model assumptions.
 
Synthetic FIR maps at $70 \mu$m,
$100 \mu$m, and $160 \mu$m (Figures \ref{pic:RT} \&
\ref{pic:RT2}) show very different FIR surface brightness
properties for varying efficiencies of
the thermal stellar feedback. We find that, for a 
rather inefficient stellar feedback formalism with $0.01 \lesssim q_{\rm
  EoS} \lesssim 0.1$, the simulated maps are in good agreement
with the spatially resolved FIR observations. This
quantitatively supports results from a simple conversion of the
  simulated local star formation to molecular hydrogen surface
  densities that compares well with the observed integrated CO
intensity map (Figure \ref{pic:H2}). We propose that the
reason for this is that the simulations employing low feedback
efficiencies encompass a sweet spot between a moderate previous gas
consumption history together with a higher susceptibility to
induced star formation after the second close encounter caused by
the weak thermal support against local gas collapse.

Probing synthetic SEDs in the FIR using different gas-to-dust
ratios, we find that a low gas-to-dust ratio of 62:1 yields
spectral shapes of the simulated SEDs that are in good agreement
with a simple modified black body spectrum fitted to the observed data
points; however, with significantly lower flux
densities in the simulated SEDs (Figure \ref{pic:SED}). For simulations
K10-Q1.0, K10-Q0.1, and K10-Q0.01 with gas-to-dust ratios of $62:1$,
the average flux decrements are $43.9 \%$, $18.6\%$, and $31.7\%$
within $50 - 280 \mu$m. The synthetic far-infrared SED of the best-fitting
simulation, K10-Q0.1 (i.e. with $q_{\rm EoS} = 0.1$),
peaks at a value of $\sim81 \Jy$ at $116 \mu$m, similar
to the peak value of $97.4 \Jy$ of the fitted modified black body spectrum
at $118 \mu$m. A comparison with the observed $L_{250}$-SFR relation
\citep[][c.f. formulae on page 13; their Figure
9]{2013MNRAS.429.2407H} reveals that the SFRs estimated from our
modelled flux densities at $250 \mu$m seem to be systematically lower
(by an average of $0.4$~dex) than
the SFRs in the hydrodynamic simulations. About half of the simulations lie
within the observed uncertainty in the SFRs, which is given as $\pm  
0.5$~dex \citep[][their Figure 9]{2013MNRAS.429.2407H}.
Again we find that the agreement is best for simulations with
low gas-to-dust ratios of $62:1$.

Is the seemingly favoured low gas-to-dust ratio of 62:1 realistic? 
Assuming all the metals to be locked up in the dust grains and none
left in the gas phase, this value corresponds to a lower 
limit in the metallicity being slightly sub-solar ($Z \sim 0.8\,
Z_\odot$). This is well within estimates for 
individual young star clusters in the Antennae having metallicities of
roughly solar \citep[$0.5 < Z/Z_\odot < 1.5$, see Figure 11 
in][]{BastianEtAl2009ApJ...701..607B}. Metal abundances determined in
the hot ISM phase are found to vary widely from sub-solar ($\sim 0.2
Z_\odot$) to highly super-solar $\sim20 Z_\odot$
\citep{FabbianoEtAl2004ApJ, 2006ApJS..162..113B}. So a lower
metallicity limit of $Z \ge 0.8\, Z_\odot$ is not ruled out by the
observations.

On the other hand, we can try and estimate the observed gas-to-dust
ratio. These estimates depend critically on the adopted CO-to-\H2
conversion factor. If we identify the total gas mass in 
the disks as $M_{\rm gas}^{\rm disk} \approx M_{\HI}^{\rm disk} +
M_{\H2}$ ($M_{\H2} = M_{\H2}^{\rm disk}$, since there is no \H2
detected in the tidal arms), we obtain 
a gas-to-dust ratio of $\mathrm{G/D} \sim 670$ for $\alpha_{\rm CO} = 4.78 \Msun
\pc^{-2}\, (K\, \kms)^{-1}$ \citep{GaoEtAl2001ApJ}, and $\mathrm{G/D}
\sim 135$ for $\alpha_{\rm CO} = 0.8 \Msun \pc^{-2}\, (K\, \kms)^{-1}$
as frequently adopted in starbursts \citep{1998ApJ...507..615D} and
consistent with the latest analysis \citep{2013arXiv1301.3498B} and
values obtained in the nucleus of NGC4038 and the overlap regions in
the Antennae \citep{2003ApJ...588..243Z}. Here we have used $M_{\H2} =
\alpha_{\rm CO}\cdot L_{\rm CO}$ with $L_{\rm CO}$ taken from
\citet{GaoEtAl2001ApJ}, $M_{\rm HI}^{\rm disk}$ from
\citet{HibbardEtAl2001AJ}, and all quantities scaled to the same
distance. While the uncertainty is large depending on our choice for
$\alpha_{\rm CO}$, these values seem to favour a higher gas-to-dust
ratio than the one we have chosen to match the observed FIR SED.  

What else could influence the simulated SED shape? Changes in the
assumed distance used to match the observations would 
result only in changes of the normalization in the simulated fluxes,
e.g. a distance change of $\sim 10 \%$ changes the fluxes by  
$\sim 20\%$. They would, however, have no influence on the
SED {\it shapes}. Higher dust masses due to higher initial gas
fractions or due to cooling and subsequent accretion from a hot
galactic halo \citep{2011MNRAS.415.3750M}, however, might lead to
redder simulated SEDs by increasing the total dust mass at any
given time throughout the merger process, without invoking low
gas-to-dust ratios. Accordingly higher SFRs might counteract this
trend, most likely, however, to a much smaller extent \citep[see
e.g.][]{2011ApJ...743..159H}. We plan to test this hypothesis in
future hydrodynamic simulations including the effects of metal
evolution and the presence of galactic hot haloes.

The specific form of the stellar initial mass function (IMF) also
affects the simulated SEDs. The star formation and feedback model
we use \citep{Springel&Hernquist2003MNRAS} intrinsically assumes a
Salpeter IMF, e.g. for computing the supernova rates and hence the
feedback energy and gas return fractions to the surrounding ISM,
producing only about half of the luminous stars with respect to a 
\citet{2003PASP..115..763C} IMF. For consistency, we also use a
Salpeter IMF to compute the stellar emission in the RT
calculations. Choosing a Chabrier IMF to test the
impact of the IMF in the RT calculations, we found that a lower
mass-to-light ratio (by $\sim 0.1-0.2$~dex) in this case leads to an
increase in the FIR emission by a similar amount ($20-50
\%$). Furthermore, the higher relative fraction of massive young stars
leads to higher dust temperatures, with the emission peaking at shorter
wavelengths and shifting the SEDs towards the blue. As a result, an
even lower gas-to-dust ratio would be required to match the observed
SEDs in Figure \ref{pic:SED}. Adopting a Chabrier IMF in the star
formation and feedback model would probably yield a more vigorous
stellar feedback due to the higher fraction in massive stars. The
consequences of these changes, however, are less obvious, requiring
further self-consistent simulations.

The radiative transfer calculation we use here assumes that the dust grains are
in thermal equilibrium with the local radiation field. However,
stochastic heating of very small dust grains by absorption of starlight
results in high-temperature transients, during which the absorbed energy
is re-radiated mainly at mid-infrared wavelengths. It has been shown
that this mechanism may have a significant effect on the emission at
70 $\mu$m, depending on the local stellar radiation field
\citep{1998ApJ...509..103S, 2001ApJ...554..778L}. At the FIR
wavelengths, which are the focus of this article, the emission from
large grains at the equilibrium temperature, however, is expected to
be dominant.

Another factor affecting the results from the radiative transfer
simulations is the treatment of small-scale structure in the
ISM. While we assume a smooth density distribution at the 
scale of individual cells ($\sim40$ pc in the central parts of the
galaxies), the ISM definitely shows structures on 
much smaller scales. We do not include any sub-scale modelling of
photodissociation regions (PDRs) around young stars or clumpy
molecular clouds. In the simulations discussed in this article the effect
of PDRs is likely to be relatively small, because a large fraction of
the total luminosity is from stars older than 10 Myr. On the other
hand, the non-uniform dust density distribution is known to have a
potentially significant effect on the dust temperature distribution
and infra-red emission \citep[e.g.][]{1996ApJ...463..681W,
  2005MNRAS.362..737D, 2007MNRAS.374..949S}. A 
proper treatment of the clumpy cloud structure using a sub-grid model
is to be considered in future studies.

In this paper we have shown as a proof of principle that variations
in parameters governing the stellar feedback model result in
unambiguous changes in the observable FIR properties. Without
attempting a fine-tuned match, in our adopted star formation and
feedback model a rather inefficient thermal stellar feedback
formulation ($0.01 \lesssim q_{\rm EoS}\lesssim 0.1$) seems to be 
clearly preferred both by simulated FIR maps and SEDs, similar to the
recently advocated range by \citet[][]{2010MNRAS.407.1529H}. Recently,
more realistic numerical models also include metal evolution, kinetic
and thermal stellar feedback \citep[e.g.][]{2011MNRAS.417..950H}, feedback
from active galactic nuclei \citep[e.g.][]{2012ApJ...754..125C},
effects from magnetic fields \citep[][]{2010ApJ...716.1438K} and
accretion from hot galactic gas haloes \citep{2009MNRAS.397..190S,
  2011MNRAS.415.3750M}. Tests with other codes, e.g. AMR
grid codes, might also prove useful to gauge numerical artefacts inherent
to different methods using the same initial conditions \citep[see
e.g.][for a comparison between SPH- and grid-based hydrodynamical
methods]{2007MNRAS.380..963A}.
The next step is to compare further mock observations of the
simulation results to multi-wavelength observations of the
Antennae. In this context, constraints from detailed X-ray maps are
particularly important \citep{FabbianoZezasMurray2001ApJ...554.1035F,
  FabbianoEtAl2004ApJ, 2006ApJS..162..113B, BaldiEtAl2006ApJ}. In
doing so, using the Antennae system as a ``standard ruler'', we can
further gauge the employed star formation algorithms and
simultaneously possibly gain further insights into the physics of
merger induced star formation.

\section*{Acknowledgments}
We thank the referee for useful suggestions and
comments which improved the presentation of the paper. 
This work was supported by the Science \& Technology Facilities
Council [grant number ST/J001538/1].
PHJ acknowledges the support of the Research Funds of the University of Helsinki.
TL and MJ acknowledge the support of the Academy of Finland through
grants No. 132291 and No. 250741. 
Many thanks to Christine Wilson for useful discussions and the
permission to reproduce Figure 1 from \citet{WilsonEtAl2000ApJ} in
Figure \ref{pic:H2}. PACS has been developed by a consortium of institutes led by
MPE (Germany) and including UVIE (Austria); KUL, CSL, IMEC (Belgium);
CEA, OAMP (France); MPIA (Germany); IFSI, OAP/AOT, OAA/CAISMI, LENS,
SISSA (Italy); IAC (Spain). This development has been supported by the
funding agencies BMVIT (Austria), ESA-PRODEX (Belgium),
CEA/CNES (France), DLR (Germany), ASI (Italy), and CICYT/MCYT (Spain).
The simulations in this paper were performed on the OPA cluster
supported by the Rechenzentrum der Max-Planck-Gesellschaft in Garching.
%



\bsp

\label{lastpage}

\end{document}